\documentclass[a4paper,10pt]{article}
\usepackage{amsmath}
\usepackage{amsfonts}
\usepackage{amssymb}
\usepackage{amsopn}
\usepackage{theorem}
\usepackage{latexsym}
\usepackage{graphicx}
\usepackage{mathrsfs}
\usepackage{psfrag}
\usepackage{url}
\usepackage{algorithmic}
\usepackage{algorithm}
\usepackage{url}

\newtheorem{result}{\ }[section]
\theoremstyle{changebreak}                
\newtheorem{thm}[result]{Theorem}

\newtheorem{lem}[result]{Lemma}
\newtheorem{cor}[result]{Corollary}

\newenvironment{proof}
 {{\sl Proof.}\hspace*{1 ex}}%
 {{\nopagebreak\hspace*{\fill}$\Box$\par\vspace{12pt}}}
\newcommand{\transpose}[1]{{#1}^{\top}}

\begin{document}

\begin{center} 

{\LARGE The Discretizable Molecular Distance Geometry Problem}
\par \bigskip
{\sc Carlile Lavor${}^1$, Leo Liberti${}^2$, Nelson Maculan${}^3$}
\par \bigskip
\begin{minipage}{15cm}
\begin{flushleft}
{\small
\begin{itemize}
\item[${}^1$] {\it Department of Applied Mathematics (IMECC-UNICAMP), 
     State University of Campinas, \\ CP 6065, 13081-970, Campinas-SP,
     Brazil} \\ E-mail:\url{clavor@ime.unicamp.br}
\item[${}^2$] {\it LIX, \'Ecole Polytechnique, F-91128 Palaiseau,
  France} \\ E-mail:\url{liberti@lix.polytechnique.fr}
\item[${}^3$] {\it COPPE -- Systems Engineering, Federal University of Rio de
     Janeiro, \\ P.O. Box 68511, 21941-972 Rio de Janeiro, Brazil.} \\
     E-mail:\url{maculan@cos.ufrj.br}
\end{itemize}
}
\end{flushleft}
\end{minipage}
\par \medskip \today
\end{center}
\par \bigskip

\begin{abstract}
Given a weighted undirected graph $G=(V,E,d)$, the Molecular Distance
Geometry Problem (MDGP) is that of finding a function $x:G\rightarrow
\mathbb{R}^{3}$, where $||x(u)-x(v)||=d(u,v)$ for each $\{u,v\}\in
E$. We show that under a few assumptions usually satisfied in
proteins, the MDGP can be formulated as a search in a discrete
space. We call this MDGP subclass the Discretizable MDGP (DMDGP). We
show that the DMDGP is \textbf{NP}-complete and we propose an
algorithm, called Branch-and-Prune (BP), which solves the DMDGP
exactly. The BP algorithm performs exceptionally well in terms of
solution accuracy and can find all solutions to any DMDGP instance.
We successfully test the BP algorithm on several randomly generated
instances. 
\end{abstract}

\thispagestyle{plain} 

\section{Introduction}
\label{intro}
It is well known that the role and function of a molecule is
determined by both its chemical structure (the atoms that compose it
and the way they bond) and its three-dimensional structure. Supposing
the chemical structure is known, finding the conformation of the atoms
in $\mathbb{R}^{3}$ is usually tackled by a mixture of chemical
analysis and mathematical methods.  Some insight as to the molecular
spatial conformation can be gained by employing Nuclear Magnetic
Resonance (NMR) techniques, which are able to give a measure of the
distance between (but not of the positions of) pairs of atoms closer
than around 5\AA . The problem of finding the atomic positions given a
subset of atomic distances can be formalized as follows.
\newline

\begin{quote}
\textsc{Molecular Distance Geometry Problem} (MDGP): given a weighted
undirected graph $G=(V,E,d)$, is there a function $x:G\rightarrow
\mathbb{R} ^{3}$ such that $||x(u)-x(v)||=d(u,v)$ for each $\{u,v\}\in
E$? \\
\end{quote}

The atoms are represented by the set of vertices $V$, the atomic positions
by $x(v)$, for $v\in V$, and the inter-atomic distance between $u$ and $v$
is given by $d(u,v)$, for $\{u,v\}\in E$. This problem has been shown to be 
\textbf{NP}-complete via a reduction from \textsc{Subset-Sum} \cite{s79},
although the problem is solvable in linear time when all the inter-atomic
distances are known \cite{dw02}. The MDGP is usually formulated as a
continuous nonconvex optimization problem: 
\begin{equation}
\min_{x}g(x)=\sum_{\{u,v\}\in E}(||x(u)-x(v)||^{2}-d(u,v)^{2})^{2}.
\label{continuous}
\end{equation}%
Obviously, $x$ solves the problem if and only if $g(x)=0$.

In practice the MDGP is usually solved via continuous optimization
methods. In \cite{h95}, the molecule is decomposed into clusters; each
cluster's 3D structure is determined independently of the others, and
then the clusters are recombined. In \cite{mw97,mw99}, a Gaussian
smoothing of (\ref{continuous}) is derived in a closed analytical form
depending on a smoothing parameter $\lambda $. The proposed algorithm
is called Global Continuation Algorithm (GCA): the smoothed problem is
locally solved for iteratively increasing values of $\lambda$ (this
brings the smoothed problem closer and closer to the original
problem), each local solution process starting from the solution of
the previous smoothing. In \cite{an,antao}, the MDGP is formulated as
a D.C. (difference of convex functions) programming problems and
solved using a variant of the D.C. Algorithm (DCA). In \cite
{lln1,doublevns}, two different Variable Neighborhood Search-based
algorithms are proposed. One of the most stringent limitations of all
these algorithms is the solution accuracy. Because there exist many
different spatial conformations having objective function values very
near zero, it is important that the optimal solution should have an
objective value as close to zero as possible. Continuous optimization
methods, by the very limitations of floating point arithmetics, are
not well suited to produce extremely accurate values. Two completely
different approaches to solving the MDGP are given in
\cite{grovermdgp} (based on quantum computation) and \cite{wang05}
(based on algebraic geometry).

A protein consists of a main backbone (a chain of atoms) and several
\textquotedblleft dangling\textquotedblright\ side chains. The NMR
technique can of course be applied to proteins in particular, and
indeed many of the algorithms to solve the general MDGP have been
tested on proteins. However, proteins have a particular structure
which makes it possible to formulate the MDGP applied to protein
backbones as a discrete search problem: this has an enormous impact on
the solution accuracy, as floating point arithmetics calculations are
fewer than with continuous search methods. We formalize this by
introducing the Discretizable Molecular Distance Geometry Problem
(DMDGP), which consists of a certain subset of MDGP instances (to
which most protein backbones belong) for which a discrete formulation
can be supplied. The determination of the spatial position of the side
chains is called the {\sc Side Chain Placement Problem} (SCPP), and
its discrete version is known to be {\bf NP}-complete
\cite{s05,s06}. Although in this paper we only consider the
determination of the protein backbone, it is clear that given a set of
likely backbones, some of them can be discarded if the resulting SCPP
instance turns out to be infeasible. In this sense, the DMDGP and the
SCPP are largely complementary. \\

\begin{quote}
{\sc Discretizable Molecular Distance Geometry Problem} (DMDGP): given
a weighted undirected graph $G=(V,E,d)$ such that there exists an
ordering $v_{1},\ldots, v_{n}\in V$ satisfying the following
requirements:
\begin{enumerate}
\item $E$ contains all cliques on quadruplets of consecutive vertices:
$\forall i\in\{4,\ldots,n\}\; \forall j,k\in\{i-3,\ldots,i\} \;
(\{j,k\}\in E)$; \label{assumption1}
\item the following strict triangular inequality holds:
$d(v_{i-1},v_{i+1})<d(v_{i-1},v_{i})+d(v_{i},v_{i+1})$, for
$i=2,\ldots,n-1$, \label{assumption2}
\end{enumerate}
is there a function $x:G\rightarrow \mathbb{R} ^{3}$ such that
$||x(u)-x(v)||=d(u,v)$ for each $\{u,v\}\in E$? \\
\end{quote}

\begin{figure}[!ht]
\psfrag{i}{$i-3$}
\psfrag{i+1}{$i-2$}
\psfrag{i+2}{$i-1$}
\psfrag{i+3}{$i$}
\psfrag{ri1}{$d_{i-3,i-2}$}
\psfrag{ri2}{$d_{i-2,i-1}$}
\psfrag{ri3}{$d_{i-1,i}$}
\psfrag{dii2}{$d_{i-3,i-1}$}
\psfrag{di1i3}{$d_{i-2,i}$}
\psfrag{ti1}{$\theta_{i-3,i-1}$}
\psfrag{ti2}{$\theta_{i-2,i}$}
\psfrag{wi3}{$\omega_{i-3,i}$}
\begin{center}
\fbox{\includegraphics[width=12cm]{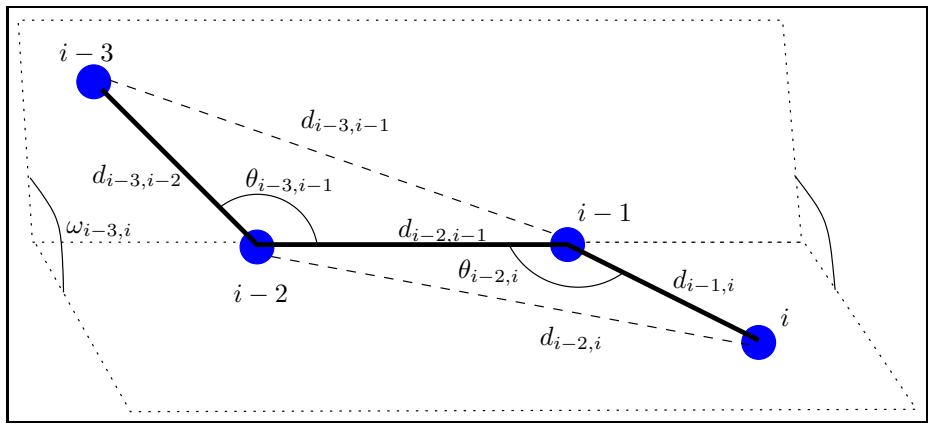}}
\end{center}
\caption{Definitions of bond lengths, bond angles and torsion angles.}
\label{figdefn}
\end{figure}

The distances $d(v_{i-1},v_{i})$ are called \textit{bond lengths}, for
$i=2,\ldots, n$, and the angles $\theta _{i-2,i}$ between atoms
$v_{i-2},v_{i-1},v_{i}$ are called \textit{bond angles}, for
$i=3,\ldots, n$ (see Fig.~\ref{figdefn}). The ordering on $V$ is
called the {\it backbone ordering}. Furthermore, we partition $E$ in
two sets $H$ and $F$ such that $H=\{\{i,j\}\in E\;|\; |i-j|\le 4\}$
and $F=E\smallsetminus H$. \label{EFH}

In practice, Assumption \ref{assumption1} requires that the bond
lengths and angles, as well as the distances between atoms separated
by three consecutive bond lengths are known. The distances between
atoms separated by two consecutive bond lengths may of course be
trivially computed from the bond lengths and angles. Assumption
\ref{assumption2} says that no bond angle may be a multiple of $\pi$.
Assumption \ref{assumption1} is applicable to many proteins as NMR is
able to compute distances of atoms which are close together, and
groups of four consecutive atoms in the backbone ordering are usually
closer than the threshold value of 5\AA\
\cite{c93,schlick}. Assumption \ref{assumption2} is also applicable to
proteins as, to the best of our knowledge, no protein has bond angles
of exactly $\pi$. In any case, the probability measure of a protein
having a bond angle of exactly $\pi$ is zero.

We propose an algorithm based on the discrete formulation, called
Branch-and-Prune (BP), to solve the DMDGP exactly. The BP algorithm is
several orders of magnitude more accurate than other existing
algorithms, and usually also much faster. Moreover, other algorithms
targeting the MDGP address the question of experimental errors by
introducing distance bounds. As it turns out, NMR not only produces
systematic measurement errors (which may be dealt with by introducing
distance bounds) but also, more importantly, a non-negligible quantity
of completely wrong measures \cite{bkl99}. The BP algorithm is able to
account for this type of error, and allows for a certain percentage of
distances to be completely wrong.

In Section~\ref{discretization}, we derive the discrete formulation of
the DMDGP. In Section~\ref{complexity}, we prove that the DMDGP is
\textbf{NP}-complete. In Section~\ref{bp}, we discuss the BP algorithm
to solve the DMDGP to optimality; Section~\ref{correction} shows how
the BP algorithm deals with the two main types of NMR error
measurements. In Section~\ref{compres}, we show computational results
on some randomly generated instances. Section~\ref{conclusion} concludes the
paper.

\section{Discrete formulation of the MDGP}
\label{discretization}
In what follows, we will restrict our attention to the
DMDGP. Notationwise, we indicate $x(v_{i})$ by $x_{i}$ and
$d(v_{i},v_{j})$ by $d_{i,j}$. For all $i\in V$, the neighbourhood
$\delta(i)$ of $i$ is the set $\{j\in V\;|\;\{j,i\}\in E\}$ of
vertices adjacent to $i$. With respect to the order $<$ on $V$ given
by the DMDGP definition, we let
$\bar{\delta}(i)=\{j\in\delta(i)\;|\;j<i\}$ for all $i\in V$.

\begin{figure}[!ht]
  \psfrag{i}{$i-3$}
  \psfrag{i1}{$i-2$}
  \psfrag{i2}{$i-1$}
  \psfrag{i+3}{$i$}
  \psfrag{(i+3)'}{$i'$}
  \psfrag{ti1}{$\theta_{i-3,i-1}$}
  \psfrag{ti2}{$\theta_{i-2,i}$}
  \psfrag{dii3}{$d_{i-3,i}$}
  \psfrag{dii3'}{$d_{i-3,i}$}
  \begin{center}
  \fbox{\includegraphics[width=12cm]{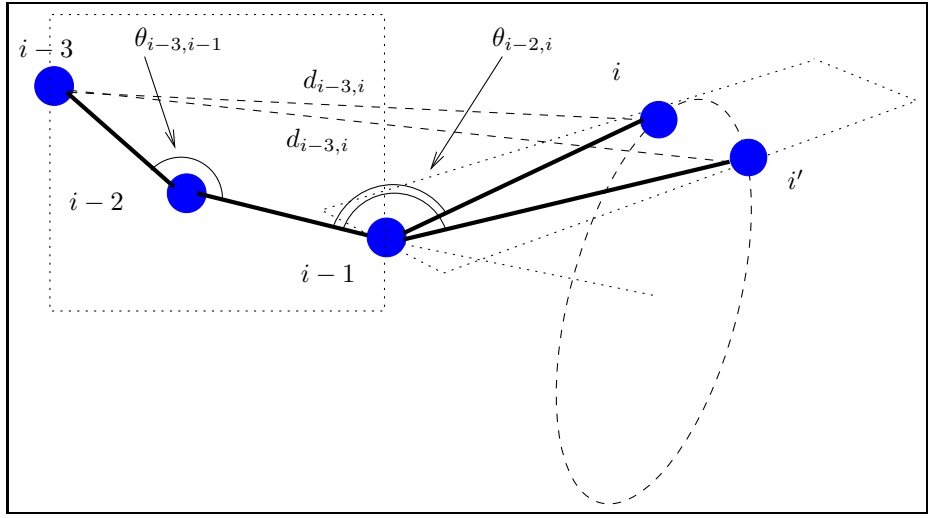}}
  \end{center}
  \caption{Discretization of the problem. The atom $i$ can only be in
  the two shown positions ($i$ and $i'$) in order to be feasible with
  the distance $d_{i-3,i}$.}
  \label{figdiscr}
\end{figure}

In order to describe a molecule with $n$ atoms, in addition to the
bond lengths $d_{i-1,i}$, for $i=2,\ldots,n$, and the bond angles
$\theta_{i-2,i}$, for $i=3,\ldots,n$, we also have to consider the
{\it torsion angles} $\omega _{i-3,i}$, for $i=4,\ldots,n$, which are
the angles between the normals through the planes defined by the atoms
$i-3,i-2,i-1$ and $i-2,i-1,i$ (see Fig.~\ref{figdefn}). However, in
most molecular conformation calculations, all the bond lengths and
bond angles are assumed to be known {\it a priori}. Thus, the first
three atoms of the molecule can be fixed and the fourth atom can be
determined by the torsion angle $\omega _{1,4}$ (see
Fig.~\ref{figdiscr}). The fifth atom can be determined by the torsion
angles $\omega_{1,4}$ and $\omega_{2,5}$, and so on.

The geometrical intuition behind the discrete formulation is that the
$i$-th atom lies on the intersection of three spheres centered at
atoms $i-3,i-2,i-1$ and of radii $d_{i-3,i},d_{i-2,i},d_{i-1,i}$
respectively. By Assumption \ref{assumption2} and the fact that no two
atoms can ever take the same position in space, the intersection of
the three spheres defines at most two points (labeled $i$ and $i'$ in
Fig.~\ref{figdiscr}). This allows us to express the position of the
$i$-th atom in terms of the preceding three, giving us $2^{n-3}$
possible molecules. Of course some of these will be infeasible with
respect to the distances in $F$ (i.e.~distance between atoms which are
further apart than 4 units in the backbone ordering), as well as with
respect to other constraints (see Sec.~\ref{bp}).

It is known that \cite{prw96}, given all the bond lengths
$d_{1,2},\ldots,d_{n-1,n}$, bond angles $\theta _{13},\ldots
,\theta_{n-2,n}$, and torsion angles $\omega_{1,4},\ldots
,\omega_{n-3,n}$ of a molecule with $n$ atoms, the Cartesian
coordinates $(x_{i_{1}},x_{i_{2}},x_{i_{3}})$ for each atom $i$ in the
molecule can be obtained using the following formulae:
\begin{equation*}
\left[ 
\begin{array}{r}
x_{i_{1}} \\ 
x_{i_{2}} \\ 
x_{i_{3}} \\ 
1
\end{array}
\right] =B_{1}B_{2}\cdots B_{i}\left[ 
\begin{array}{r}
0 \\ 
0 \\ 
0 \\ 
1
\end{array}
\right] ,\text{ }\forall i=1,\ldots ,n,
\end{equation*}
where 
\begin{equation}
B_{1}=\left[ 
\begin{array}{rrrr}
1 & 0 & 0 & 0 \\ 
0 & 1 & 0 & 0 \\ 
0 & 0 & 1 & 0 \\ 
0 & 0 & 0 & 1
\end{array}
\right] ,\text{ \ }B_{2}=\left[ 
\begin{array}{rrrr}
-1 & 0 & 0 & -d_{1,2} \\ 
0 & 1 & 0 & 0 \\ 
0 & 0 & -1 & 0 \\ 
0 & 0 & 0 & 1
\end{array}
\right] ,  \label{b1b2}
\end{equation}
\begin{equation*}
B_{3}=\left[ 
\begin{array}{rrrr}
-\cos \theta _{1,3} & -\sin \theta _{1,3} & 0 & -d_{2,3}\cos \theta _{1,3}
\\ 
\sin \theta _{1,3} & -\cos \theta _{1,3} & 0 & d_{2,3}\sin \theta _{1,3} \\ 
0 & 0 & 1 & 0 \\ 
0 & 0 & 0 & 1
\end{array}
\right] ,
\end{equation*}
and 
\begin{equation}
B_{i}=\left[ 
\begin{array}{rrrr}
-\cos \theta _{i-2,i} & -\sin \theta _{i-2,i} & 0 & -d_{i-1,i}\cos \theta
_{i-2,i} \\ 
\sin \theta _{i-2,i}\cos \omega _{i-3,i} & -\cos \theta _{i-2,i}\cos \omega
_{i-3,i} & -\sin \omega _{i-3,i} & d_{i-1,i}\sin \theta _{i-2,i}\cos \omega
_{i-3,i} \\ 
\sin \theta _{i-2,i}\sin \omega _{i-3,i} & -\cos \theta _{i-2,i}\sin \omega
_{i-3,i} & \cos \omega _{i-3,i} & d_{i-1,i}\sin \theta _{i-2,i}\sin \omega
_{i-3,i} \\ 
0 & 0 & 0 & 1
\end{array}
\right] ,  \label{bi}
\end{equation}
for $i=4,...,n$.

Since all the bond lengths and bond angles are assumed to be given in
the instance, the Cartesian coordinates of all atoms of a molecule can
be completely determined by using the values of $\cos \omega _{i-3,i}$
and $\sin \omega _{i-3,i}$, for $i=4,...,n$.

In order to state that the DMDGP can be formulated as a search in a discrete
space, we need the following lemma.

\begin{lem}
For instances of the DMDGP\ class, for all $i=4,...,n$, the value of $\cos
\omega_{i-3,i}$ can be computed $O(1)$.
\label{lemma21}
\end{lem}

\begin{proof}
This follows because for every four consecutive atoms
$x_{i-3},x_{i-2},x_{i-1},x_{i}$, the cosine of the torsion angle
$\omega _{i-3,i}$, for $i=4,...,n$, is given by
\begin{equation}
\cos
\omega_{i-3,i}=\frac{d_{i-3,i-2}^{2}+d_{i-2,i}^{2}-2d_{i-3,i-2}d_{i-2,i}
\cos \theta_{i-2,i}\cos
\theta_{i-1,i+1}-d_{i-3,i}^{2}}{2d_{i-3,i-2}d_{i-2,i}\sin
\theta_{i-2,i}\sin \theta_{i-1,i+1}}, \label{pog}
\end{equation}
which is just a rearrangement of the cosine law for torsion angles \cite{p87}
(p. 278), and all the values in the expression (\ref{pog}) are given in the
instance. We note in passing that in order for the above reasoning to hold,
we obviously need the denominator of (\ref{pog}) to be nonzero.
\end{proof}

\begin{thm}
Given a weighted undirected graph $G=(V,E,d)$ associated to an instance of
the\ DMDGP, the number of functions $x:G\rightarrow \mathbb{R}^{3}$ such
that $||x(u)-x(v)||=d(u,v)$ for each $\{u,v\}\in E$ is finite, up to
orthogonality transformations.
\end{thm}

\begin{proof}
The proof is by induction. For a molecule with 4 atoms, we can use the bond
lengths $d_{1,2},d_{2,3}$ and the bond angle $\theta _{1,3}$, in order to
determine the matrices $B_{2}$ and $B_{3}$, defined in (\ref{b1b2}), and
obtain 
\begin{eqnarray*}
x_{1} &=&\left( 
\begin{array}{c}
0 \\ 
0 \\ 
0
\end{array}
\right) , \\
x_{2} &=&\left( 
\begin{array}{c}
-d_{1,2} \\ 
0 \\ 
0
\end{array}
\right) , \\
x_{3} &=&\left( 
\begin{array}{c}
-d_{1,2}+d_{2,3}\cos \theta _{1,3} \\ 
d_{2,3}\sin \theta _{1,3} \\ 
0
\end{array}
\right),
\end{eqnarray*}
fixing the first three atoms of the molecule. Since we also know the
distance $d_{1,4}$, by Lemma \ref{lemma21} we can obtain the value of
$\cos \omega_{1,4}$. Thus, the sine of the torsion angle $\omega
_{1,4}$ can have only two possible values: $\sin \omega _{1,4}=\pm
\sqrt{1-\cos ^{2}\omega _{1,4}}$. Consequently, by (\ref{bi}), we
obtain only two possible positions $x_{4},x_{4}^{\prime }$ for the
fourth atom of the molecule, given by
\begin{eqnarray*}
x_{4} &=&\left[ 
\begin{array}{r}
-d_{1,2}+d_{2,3}\cos \theta _{1,3}-d_{3,4}\cos \theta _{1,3}\cos \theta
_{2,4}+d_{3,4}\sin \theta _{1,3}\sin \theta _{2,4}\cos \omega _{1,4} \\ 
d_{2,3}\sin \theta _{1,3}-d_{3,4}\sin \theta _{1,3}\cos \theta
_{2,4}-d_{3,4}\cos \theta _{1,3}\sin \theta _{2,4}\cos \omega _{1,4} \\ 
d_{3,4}\sin \theta _{2,4}\left( \sqrt{1-\cos ^{2}\omega _{1,4}}\right) 
\end{array}
\right] , \\
x_{4}^{\prime } &=&\left[ 
\begin{array}{r}
-d_{1,2}+d_{2,3}\cos \theta _{1,3}-d_{3,4}\cos \theta _{1,3}\cos \theta
_{2,4}+d_{3,4}\sin \theta _{1,3}\sin \theta _{2,4}\cos \omega _{1,4} \\ 
d_{2,3}\sin \theta _{1,3}-d_{3,4}\sin \theta _{1,3}\cos \theta
_{2,4}-d_{3,4}\cos \theta _{1,3}\sin \theta _{2,4}\cos \omega _{1,4} \\ 
d_{3,4}\sin \theta _{2,4}\left( -\sqrt{1-\cos ^{2}\omega _{1,4}}\right) 
\end{array}
\right].
\end{eqnarray*}

Now suppose that for $i\geq 4$ atoms, we have a finite number of
solutions to the DMDGP instance. Adding one more atom in the molecule
and using Lemma \ref{lemma21} again, we can obtain the value of $\cos
\omega _{i-2,i+1}$. From each solution of the molecule with $i$ atoms,
at most two new solutions can be obtained by using $\sin \omega
_{i-2,i+1}=\pm \sqrt{1-\cos ^{2}\omega _{i-2,i+1}}$ in matrix
$B_{i+1}$, given in (\ref{bi}). This concludes the proof.
\end{proof}

An immediate corollary is given below.

\begin{cor}
For an instance of the\ DMDGP class with $n\geq 4$ atoms, there are at
most $2^{n-3}$ possible solutions.
\end{cor}

Note that each possible solution of the DMDGP is defined by a sequence of
torsion angles $\omega _{1,4},\ldots ,\omega _{n-3,n}$. By using the
matrices $B_{i}$ (\ref{bi}), this sequence can be converted into another one
of Cartesian coordinates $x=(x_{1},\ldots ,x_{n})\in \mathbb{R}^{3n}$ and,
using the objective function $g$ defined in (\ref{continuous}), a solution
can be identified simply by testing if $g(x)=0$.

\subsection{Solution symmetry}
\label{symmetry}
In this section, we show that there is a solution symmetry around the
plane defined by the first three atoms; more precisely, any solution
on one side of this plane gives rise to a symmetrical solution on the
other side. This allows us to reduce computational costs by
half. First, we need two lemmata. \\

\begin{lem}
Let the matrix $Q_{i}$ be defined by
\begin{equation*}
Q_{i}=B_{4}\cdots B_{i},
\end{equation*}
for $i=4,...,n$, where its elements are denoted by
\begin{equation*}
Q_{i}=\left[ 
\begin{array}{rrrr}
q_{11}^{i} & q_{12}^{i} & q_{13}^{i} & q_{14}^{i} \\ 
q_{21}^{i} & q_{22}^{i} & q_{23}^{i} & q_{24}^{i} \\ 
q_{31}^{i} & q_{32}^{i} & q_{33}^{i} & q_{34}^{i} \\ 
0 & 0 & 0 & 1
\end{array}
\right].
\end{equation*}
If we invert the sign of $\sin \omega _{i-3,i}$ in all the matrices
$B_{i}$ (\ref{bi}), for $i=4,...,n$, and denote the new matrices
obtained by $B_{i}^{\prime }$, then the elements of the matrix
$Q_{i}^{\prime}$, defined by
\begin{equation*}
Q_{i}^{\prime }=B_{4}^{\prime }\cdots B_{i}^{\prime },
\end{equation*}
is given by
\begin{equation*}
Q_{i}^{\prime }=\left[ 
\begin{array}{rrrr}
q_{11}^{i} & q_{12}^{i} & -q_{13}^{i} & q_{14}^{i} \\ 
q_{21}^{i} & q_{22}^{i} & -q_{23}^{i} & q_{24}^{i} \\ 
-q_{31}^{i} & -q_{32}^{i} & q_{33}^{i} & -q_{34}^{i} \\ 
0 & 0 & 0 & 1
\end{array}
\right],
\end{equation*}
for $i=4,...,n$.
\end{lem}

\begin{proof}
The proof is by induction. For $n=4$, we obtain: 
\begin{equation*}
Q_{4}=\left[ 
\begin{array}{rrrr}
-\cos \theta _{2,4} & -\sin \theta _{2,4} & 0 & -d_{3,4}\cos \theta _{2,4}
\\ 
\sin \theta _{2,4}\cos \omega _{1,4} & -\cos \theta _{2,4}\cos \omega _{1,4}
& -\sin \omega _{1,4} & d_{3,4}\sin \theta _{2,4}\cos \omega _{1,4} \\ 
\sin \theta _{2,4}\sin \omega _{1,4} & -\cos \theta _{2,4}\sin \omega _{1,4}
& \cos \omega _{1,4} & d_{3,4}\sin \theta _{2,4}\sin \omega _{1,4} \\ 
0 & 0 & 0 & 1
\end{array}
\right]
\end{equation*}
and 
\begin{equation*}
Q_{4}^{\prime }=\left[ 
\begin{array}{rrrr}
-\cos \theta _{2,4} & -\sin \theta _{2,4} & 0 & -d_{3,4}\cos \theta _{2,4}
\\ 
\sin \theta _{2,4}\cos \omega _{1,4} & -\cos \theta _{2,4}\cos \omega _{1,4}
& -\left( -\sin \omega _{1,4}\right) & d_{3,4}\sin \theta _{2,4}\cos \omega
_{1,4} \\ 
\sin \theta _{2,4}\left( -\sin \omega _{1,4}\right) & -\cos \theta
_{2,4}\left( -\sin \omega _{1,4}\right) & \cos \omega _{1,4} & d_{3,4}\sin
\theta _{2,4}\left( -\sin \omega _{1,4}\right) \\ 
0 & 0 & 0 & 1
\end{array}
\right].
\end{equation*}

Suppose now that the assertion is valid for $n=i-1$. Rewritting $Q_{i}$, we
get 
\begin{eqnarray*}
Q_{i} &=&(B_{4}\cdots B_{i-1})B_{i} \\
&=&Q_{i-1}B_{i},
\end{eqnarray*}
where the elements of $Q_{i-1}$ are denoted by 
\begin{equation*}
Q_{i-1}=\left[ 
\begin{array}{rrrr}
q_{11}^{i-1} & q_{12}^{i-1} & q_{13}^{i-1} & q_{14}^{i-1} \\ 
q_{21}^{i-1} & q_{22}^{i-1} & q_{23}^{i-1} & q_{24}^{i-1} \\ 
q_{31}^{i-1} & q_{32}^{i-1} & q_{33}^{i-1} & q_{34}^{i-1} \\ 
0 & 0 & 0 & 1
\end{array}
\right] 
\end{equation*}
and 
\begin{equation*}
B_{i}=\left[ 
\begin{array}{rrrr}
-\cos \theta _{i-2,i} & -\sin \theta _{i-2,i} & 0 & -d_{i-1,i}\cos \theta
_{i-2,i} \\ 
\sin \theta _{i-2,i}\cos \omega _{i-3,i} & -\cos \theta _{i-2,i}\cos \omega
_{i-3,i} & -\sin \omega _{i-3,i} & d_{i-1,i}\sin \theta _{i-2,i}\cos \omega
_{i-3,i} \\ 
\sin \theta _{i-2,i}\sin \omega _{i-3,i} & -\cos \theta _{i-2,i}\sin \omega
_{i-3,i} & \cos \omega _{i-3,i} & d_{i-1,i}\sin \theta _{i-2,i}\sin \omega
_{i-3,i} \\ 
0 & 0 & 0 & 1
\end{array}
\right].
\end{equation*}
Considering the product $Q_{i-1}B_{i}$, we obtain
\begin{equation*}
Q_{i-1}B_{i}=\left[ 
\begin{array}{cccc}
V & X & Y & Z \\ 
0 & 0 & 0 & 1
\end{array}
\right],
\end{equation*}
where 
\begin{eqnarray*}
V &=&\left[ 
\begin{array}{r}
q_{11}^{i-1}(-b)+q_{12}^{i-1}(cd)+q_{13}^{i-1}(ce) \\ 
q_{21}^{i-1}(-b)+q_{22}^{i-1}(cd)+q_{23}^{i-1}(ce) \\ 
q_{31}^{i-1}(-b)+q_{32}^{i-1}(cd)+q_{33}^{i-1}(ce)
\end{array}
\right] , \\
X &=&\left[ 
\begin{array}{r}
q_{11}^{i-1}(-c)+q_{12}^{i-1}(-bd)+q_{13}^{i-1}(-be) \\ 
q_{21}^{i-1}(-c)+q_{22}^{i-1}(-bd)+q_{23}^{i-1}(-be) \\ 
q_{31}^{i-1}(-c)+q_{32}^{i-1}(-bd)+q_{33}^{i-1}(-be)
\end{array}
\right] , \\
Y &=&\left[ 
\begin{array}{c}
q_{12}^{i-1}(-e)+q_{13}^{i-1}(d) \\ 
q_{22}^{i-1}(-e)+q_{23}^{i-1}(d) \\ 
q_{32}^{i-1}(-e)+q_{33}^{i-1}(d)
\end{array}
\right] , \\
Z &=&\left[ 
\begin{array}{r}
q_{11}^{i-1}(-ab)+q_{12}^{i-1}(acd)+q_{13}^{i-1}(ace)+q_{14}^{i-1} \\ 
q_{21}^{i-1}(-ab)+q_{22}^{i-1}(acd)+q_{23}^{i-1}(ace)+q_{24}^{i-1} \\ 
q_{31}^{i-1}(-ab)+q_{32}^{i-1}(acd)+q_{33}^{i-1}(ace)+q_{34}^{i-1}
\end{array}
\right],
\end{eqnarray*}
and $a=d_{i-1,i}$, $b=\cos \theta _{i-2,i}$, $c=\sin \theta _{i-2,i}$,
$d=\cos \omega _{i-3,i}$, and $e=\sin \omega _{i-3,i}$.

By induction hypothesis, we have 
\begin{equation*}
Q_{i-1}^{\prime }=\left[ 
\begin{array}{rrrr}
q_{11}^{i-1} & q_{12}^{i-1} & -q_{13}^{i-1} & q_{14}^{i-1} \\ 
q_{21}^{i-1} & q_{22}^{i-1} & -q_{23}^{i-1} & q_{24}^{i-1} \\ 
-q_{31}^{i-1} & -q_{32}^{i-1} & q_{33}^{i-1} & -q_{34}^{i-1} \\ 
0 & 0 & 0 & 1
\end{array}
\right].
\end{equation*}
Considering the product $Q_{i-1}^{\prime }B_{i}^{\prime }$, where 
\begin{equation*}
B_{i}^{\prime }=\left[ 
\begin{array}{rrrr}
-\cos \theta _{i-2,i} & -\sin \theta _{i-2,i} & 0 & -d_{i-1,i}\cos \theta
_{i-2,i} \\ 
\sin \theta _{i-2,i}\cos \omega _{i-3,i} & -\cos \theta _{i-2,i}\cos \omega
_{i-3,i} & -\left( -\sin \omega _{i-3,i}\right)  & d_{i-1,i}\sin \theta
_{i-2,i}\cos \omega _{i-3,i} \\ 
\sin \theta _{i-2,i}(-\sin \omega _{i-3,i}) & -\cos \theta _{i-2,i}(-\sin
\omega _{i-3,i}) & \cos \omega _{i-3,i} & d_{i-1,i}\sin \theta
_{i-2,i}(-\sin \omega _{i-3,i}) \\ 
0 & 0 & 0 & 1
\end{array}
\right],
\end{equation*}
we obtain
\begin{equation*}
Q_{i-1}^{\prime }B_{i}^{\prime }=\left[ 
\begin{array}{cccc}
V^{\prime } & X^{\prime } & Y^{\prime } & Z^{\prime } \\ 
0 & 0 & 0 & 1
\end{array}
\right],
\end{equation*}
where 
\begin{eqnarray*}
V^{\prime } &=&\left[ 
\begin{array}{r}
q_{11}^{i-1}(-b)+q_{12}^{i-1}(cd)-q_{13}^{i-1}(c(-e)) \\ 
q_{21}^{i-1}(-b)+q_{22}^{i-1}(cd)-q_{23}^{i-1}(c(-e)) \\ 
-q_{31}^{i-1}(-b)-q_{32}^{i-1}(cd)+q_{33}^{i-1}(c(-e))
\end{array}
\right], \\
X^{\prime } &=&\left[ 
\begin{array}{r}
q_{11}^{i-1}(-c)+q_{12}^{i-1}(-bd)-q_{13}^{i-1}(-b(-e)) \\ 
q_{21}^{i-1}(-c)+q_{22}^{i-1}(-bd)-q_{23}^{i-1}(-b(-e)) \\ 
-q_{31}^{i-1}(-c)-q_{32}^{i-1}(-bd)+q_{33}^{i-1}(-b(-e))
\end{array}
\right], \\
Y^{\prime } &=&\left[ 
\begin{array}{r}
q_{12}^{i-1}(e)-q_{13}^{i-1}(d) \\ 
q_{22}^{i-1}(e)-q_{23}^{i-1}(d) \\ 
-q_{32}^{i-1}(e)+q_{33}^{i-1}(d)
\end{array}
\right], \\
Z^{\prime } &=&\left[ 
\begin{array}{r}
q_{11}^{i-1}(-ab)+q_{12}^{i-1}(acd)-q_{13}^{i-1}(ac(-e))+q_{14}^{i-1} \\ 
q_{21}^{i-1}(-ab)+q_{22}^{i-1}(acd)-q_{23}^{i-1}(ac(-e))+q_{24}^{i-1} \\ 
-q_{31}^{i-1}(-ab)-q_{32}^{i-1}(acd)+q_{33}^{i-1}(ac(-e))-q_{34}^{i-1}
\end{array}
\right].
\end{eqnarray*}

Representing the matrix $Q_{i}$ by 
\begin{equation*}
Q_{i}=Q_{i-1}B_{i}=\left[ 
\begin{array}{rrrr}
q_{11}^{i} & q_{12}^{i} & q_{13}^{i} & q_{14}^{i} \\ 
q_{21}^{i} & q_{22}^{i} & q_{23}^{i} & q_{24}^{i} \\ 
q_{31}^{i} & q_{32}^{i} & q_{33}^{i} & q_{34}^{i} \\ 
0 & 0 & 0 & 1
\end{array}
\right]
\end{equation*}
and comparing the matrices $Q_{i-1}B_{i}$ and $Q_{i-1}^{\prime
}B_{i}^{\prime }$ given above, we conclude that 
\begin{equation*}
Q_{i}^{\prime }=Q_{i-1}^{\prime }B_{i}^{\prime }=\left[ 
\begin{array}{rrrr}
q_{11}^{i} & q_{12}^{i} & -q_{13}^{i} & q_{14}^{i} \\ 
q_{21}^{i} & q_{22}^{i} & -q_{23}^{i} & q_{24}^{i} \\ 
-q_{31}^{i} & -q_{32}^{i} & q_{33}^{i} & -q_{34}^{i} \\ 
0 & 0 & 0 & 1
\end{array}
\right].
\end{equation*}
\end{proof}

\begin{lem}
Le $x_{1},\ldots ,x_{n}\in \mathbb{R}^{3}$ be the Cartesian
coordinates associated to the torsion angles $\omega _{1,4},\ldots
,\omega _{n-3,n}$. If we invert the sign of $\sin \omega _{i-3,i}$ in
all the matrices $B_{i}$ (\ref{bi}), for $i=4,...,n$, then the new
Cartesian coordinates $x_{1}^{\prime },\ldots ,x_{n}^{\prime }\in
\mathbb{R}^{3}$ is given by
\begin{equation*}
\left[ 
\begin{array}{r}
x_{i_{1}}^{\prime } \\ 
x_{i_{2}}^{\prime } \\ 
x_{i_{3}}^{\prime }
\end{array}
\right] =\left[ 
\begin{array}{r}
x_{i_{1}} \\ 
x_{i_{2}} \\ 
-x_{i_{3}}
\end{array}
\right],
\end{equation*}
for $i=1,...,n$.
\end{lem}

\begin{proof}
For $n=1,2,3$ the assertion is clearly true. By the lemma above, we have 
\begin{equation*}
\left[ 
\begin{array}{r}
x_{i_{1}} \\ 
x_{i_{2}} \\ 
x_{i_{3}} \\ 
1
\end{array}
\right] =B_{1}B_{2}B_{3}Q_{i}\left[ 
\begin{array}{r}
0 \\ 
0 \\ 
0 \\ 
1
\end{array}
\right] =B_{1}B_{2}B_{3}\left[ 
\begin{array}{r}
q_{14}^{i} \\ 
q_{24}^{i} \\ 
q_{34}^{i} \\ 
1
\end{array}
\right]
\end{equation*}
and 
\begin{equation*}
\left[ 
\begin{array}{r}
x_{i_{1}}^{\prime } \\ 
x_{i_{2}}^{\prime } \\ 
x_{i_{3}}^{\prime } \\ 
1
\end{array}
\right] =B_{1}B_{2}B_{3}Q_{i}^{\prime }\left[ 
\begin{array}{r}
0 \\ 
0 \\ 
0 \\ 
1
\end{array}
\right] =B_{1}B_{2}B_{3}\left[ 
\begin{array}{r}
q_{14}^{i} \\ 
q_{24}^{i} \\ 
-q_{34}^{i} \\ 
1
\end{array}
\right],
\end{equation*}
for $i=4,...,n$, and calculating the product $B_{1}B_{2}B_{3}$, we obtain 
\begin{equation*}
B_{1}B_{2}B_{3}=\left[ 
\begin{array}{rrrr}
\cos \theta _{1,3} & \sin \theta _{1,3} & 0 & -d_{1,2}+d_{2,3}\cos \theta
_{1,3} \\ 
\sin \theta _{1,3} & -\cos \theta _{1,3} & 0 & d_{2,3}\sin \theta _{1,3} \\ 
0 & 0 & -1 & 0 \\ 
0 & 0 & 0 & 1
\end{array}
\right].
\end{equation*}
Thus, 
\begin{equation*}
\left[ 
\begin{array}{r}
x_{i_{1}} \\ 
x_{i_{2}} \\ 
x_{i_{3}} \\ 
1
\end{array}
\right] =B_{1}B_{2}B_{3}\left[ 
\begin{array}{r}
q_{14}^{i} \\ 
q_{24}^{i} \\ 
q_{34}^{i} \\ 
1
\end{array}
\right] =\left[ 
\begin{array}{r}
-d_{1,2}+q_{14}^{i}\cos \theta _{1,3}+q_{24}^{i}\sin \theta
_{1,3}+d_{2,3}\cos \theta _{1,3} \\ 
q_{14}^{i}\sin \theta _{1,3}-q_{24}^{i}\cos \theta _{1,3}+d_{2,3}\sin \theta
_{1,3} \\ 
-q_{34}^{i} \\ 
1
\end{array}
\right]
\end{equation*}
and 
\begin{equation*}
\left[ 
\begin{array}{r}
x_{i_{1}}^{\prime } \\ 
x_{i_{2}}^{\prime } \\ 
x_{i_{3}}^{\prime } \\ 
1
\end{array}
\right] =B_{1}B_{2}B_{3}\left[ 
\begin{array}{r}
q_{14}^{i} \\ 
q_{24}^{i} \\ 
-q_{34}^{i} \\ 
1
\end{array}
\right] =\left[ 
\begin{array}{r}
-d_{1,2}+q_{14}^{i}\cos \theta _{1,3}+q_{24}^{i}\sin
\theta_{1,3}+d_{2,3}\cos \theta _{1,3} \\ q_{14}^{i}\sin \theta
_{1,3}-q_{24}^{i}\cos \theta _{1,3}+d_{2,3}\sin \theta_{1,3} \\
q_{34}^{i} \\ 1
\end{array}
\right],
\end{equation*}
for $i=4,...,n$. That is, 
\begin{equation*}
\left[ 
\begin{array}{r}
x_{i_{1}}^{\prime } \\ 
x_{i_{2}}^{\prime } \\ 
x_{i_{3}}^{\prime }
\end{array}
\right] =\left[ 
\begin{array}{r}
x_{i_{1}} \\ 
x_{i_{2}} \\ 
-x_{i_{3}}
\end{array}
\right],
\end{equation*}
for $i=1,...,n$.
\end{proof}

Finally, we can prove the following theorem.

\begin{thm}
\label{symthm}
Consider a solution $x:G\rightarrow \mathbb{R}^{3}$ for the DMDGP,
defined by the torsion angles $\omega _{1,4},\ldots ,\omega
_{n-3,n}$. If we invert the sign of $\sin \omega _{i-3,i}$ in all the
matrices $B_{i}$ (\ref{bi}), for $i=4,...,n$, we obtain a new solution
$x^{\prime }:G\rightarrow \mathbb{R}^{3}$ for the DMDGP.
\end{thm}
\begin{proof}
Le $x_{1},\ldots ,x_{n}\in \mathbb{R}^{3}$ be the Cartesian
coordinates associated to the torsion angles $\omega _{1,4},\ldots
,\omega _{n-3,n}$, $x_{1}^{\prime },\ldots ,x_{n}^{\prime }\in
\mathbb{R}^{3}$ be the Cartesian coordinates of the new solution
obtained by inverting the sign of $\sin \omega _{i-3,i}$ in all the
matrices $B_{i}$, for $i=4,...,n$, and $R:\mathbb{R}^{3}\rightarrow
\mathbb{R}^{3}$ be the function defined by
\begin{equation*}
R(x_{i_{1}},x_{i_{2}},x_{i_{3}})=(x_{i_{1}},x_{i_{2}},-x_{i_{3}}).
\end{equation*}
Since $R$ is a unitary operator, 
\begin{equation}
||x_{i}-x_{j}||=||R(x_{i})-R(x_{j})||\text{\ \ }\forall (i,j)\in E,
\label{ref1}
\end{equation}
where $E$ is the set of pairs of atoms $(i,j)$ whose Euclidean
distances $d_{i,j}$ are known for the solution $x$. By the Lemma
$2.4.$,
\begin{equation}
||R(x_{i})-R(x_{j})||=||x_{i}^{\prime }-x_{j}^{\prime }||\text{ \ }\forall
(i,j)\in E.  \label{ref2}
\end{equation}
Since $x_{1},\ldots ,x_{n}$ is a solution for the DMDGP, 
\begin{equation*}
||x_{i}-x_{j}||=d_{i,j}\text{ \ }\forall (i,j)\in E\text{.}
\end{equation*}
Thus, by (\ref{ref1}) and (\ref{ref2}), we get 
\begin{equation*}
||x_{i}^{\prime }-x_{j}^{\prime }||=d_{i,j}\text{ \ }\forall (i,j)\in E,
\end{equation*}
stating that $x_{1}^{\prime },\ldots ,x_{n}^{\prime }$ is also a solution
for the DMDGP.
\end{proof}

\subsection{Undiscretizable instances}
\label{secundiscr}
\begin{figure}[!ht]
  \psfrag{i}{$i-3$}
  \psfrag{i+3}{$i$}
  \psfrag{thetai+1}{}
  \psfrag{(i+3)'}{$i'$}
  \psfrag{dii3}{$d_{i-3,i}$}
  \psfrag{dii3'}{$d_{i-3,i}$}
  \begin{center}
  \fbox{\includegraphics[width=7cm]{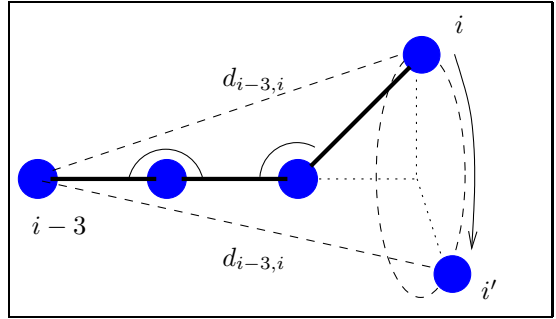}}
  \end{center}
  \caption{An instance which cannot be discretized. The $i$-th
  atom can be on any position on the circle shown without affecting
  the feasibility of the distance $d_{i-3,i}$.}
  \label{fignondiscr}
\end{figure}

As has been remarked, all DMDGP instances must obey a {\it strict}
triangular inequality. When this does not hold, there may be bond
angles with values $k\pi$ for $k\in\mathbb{Z}$. By (\ref{pog}), the
torsion angle is undefined. Since the torsion angle is the angle
between two normal vectors to given planes, it is undefined when at
least one of the planes is undefined. This is indeed possible if the
two vectors defining the plane are collinear. In other words, if a
bond angle is a multiple of $\pi$, we have the situation depicted in
Fig.~\ref{fignondiscr}, where $d_{i-3,i}$ is feasible for every
position of atom $i+3$ on the circle shown in the drawing.

Since the set $\{\pi\}$ has measure 0 in $[0,2\pi]$, the probability
that any given protein gives rise to an undiscretizable instance is
0. To the best of our knowledge, no protein has bond angles of {\it
exactly} $\pi$.

\section{Complexity}
\label{complexity}
In this section we show that the DMDGP is \textbf{NP}-complete by reducing
it from the \textsc{Subset-Sum} problem.

\begin{quote}
\textsc{Subset-Sum}. Given integers $a_{1},\ldots,a_{n}$, is there is a
partition into two sets, encoded by $s\in\{-1,+1\}^{n}$, such that each
subset has the same sum, i.e. $\sum_{i=1}^{n} s(i) a_{i}=0$?
\end{quote}

The MDGP is shown to be \textbf{NP}-complete in \cite{s79} (a helpful
sketch of the proof is given in \cite{mw97}) by reducing
\textsc{Subset-Sum} to a 1-dimensional MDGP with distance constraints
between successive atom (in an arbitrary atomic ordering) plus a
single distance constraint between the first and last atom, forcing
this distance to be zero. In the DMDGP however, we have additional
distance constraints between any pairs of atoms $1,2$ or $3$ indices
apart in the atom sequence.

\begin{thm}
The DMDGP is \textbf{NP}-complete.
\label{complexitythm}
\end{thm}
\begin{proof}
We reduce from \textsc{Subset-Sum}. Given an instance $a_{1},\ldots ,a_{n}$
of the latter, we define an instance of DMDGP on $3n+1$ points numbered $0$
to $3n$, with the following distance constraints: 
\begin{align}
d_{i,i+1}& =a_{\lfloor i/3\rfloor } & \forall i& \in \{1,...,3n-1\}
\label{eq:return} \\
d_{i,i+2}& =\sqrt{d_{i,i+1}^{2}+d_{i+1,i+2}^{2}} & \forall i& \in
\{1,...,3n-2\}  \label{eq:angle} \\
d_{i,i+3}& =\sqrt{d_{i,i+1}^{2}+d_{i+1,i+2}^{2}+d_{i+2,i+3}^{2}} & \forall
i& \in \{1,...,3n-3\}  \label{eq:xyz} \\
d_{0,3n}& =0 & &
\end{align}
Now we claim that the {\sc Subset-Sum} instance is has a solution iff
the DMDGP instance has a solution.

For the easy direction, let $s\in\{-1,+1\}^{n}$ be a solution to the
\textsc{Subset-Sum}-problem. We define the $3n+1$ points as follows:
$x_{0}=(0,0,0)$ and for every $0<i\le 3n$ with $i=3k+j$ we set
$x_{i}=x_{i-1}+s_{k}a_{k} e_{j}$, where $e_{0}=(1,0,0), e_{1}=(0,1,0)$
and $e_{2}=(0,0,1)$. By straightforward inspection this is a solution
to the DMDGP instance.

For the other direction, let us assume that the DMDGP instance has a
solution $x(v_{1}),\ldots ,x(v_{n})$. Without loss of generality we
can assume that the $x(v_{1})=(0,0,0)$, and that $x(v_{2})$ lays on
the $x$-axis. Now equation~(\ref{eq:angle}) implies that the bond
angle between $ x(v_{1}),x(v_{2}),x(v_{3})$ is $\frac{\pi
}{2}$. Again, without loss of generality assume that the second
segment is parallel to the $y$-axis. By equation~(\ref{eq:xyz}) there
are only two possibilities for $x(v_{4})$, and they force the third
bond to be parallel to the $z$-axis. The same arguments apply to all
other bonds, which shows that the bond $\beta $ between $v_{i-1}$ and
$v_{i}$ is parallel to the $(i\bmod3)$-th axis (where $x=0,y=1,z=2$).
Now give the bond $\beta $ an orientation from $v_{i-1}$ to $v_{i}$
(which can either be in the same or in the opposite direction of this
axis). We define a sign vector $s\in \{-1,+1\}^{3n}$, which encodes
these orientations. In this setting, point $3n+1$ has coordinates
$(x,y,z)$ defined by
\begin{align*}
x& =\sum_{i\bmod3=0}s_{i}a_{i} \\
y& =\sum_{i\bmod3=1}s_{i}a_{i} \\
z& =\sum_{i\bmod3=2}s_{i}a_{i}
\end{align*}
By equation~(\ref{eq:return}) we actually have $(x,y,z)=(0,0,0)$. Now
let $s^{0},s^{1},s^{2}$ be three vectors from $\{-1,+1\}^{n}$, which
are $s$ restricted to indices $i\bmod3=0$, $i\bmod3=1$ or $i\bmod3=2$
respectively.  Then any of those is a solution to the original
\textsc{Subset-Sum} problem by the previous equations.
\end{proof}

It is interesting to note that Assumption \ref{assumption1} is, in a
certain sense, the tightest possible for the problem to be
\textbf{NP}-complete.  Assumption \ref{assumption1} states that each
quadruplet of consecutive vertices in the defined order is a clique in
the distance graph. Tightening the assumption further, we might ask
whether the problem would still be \textbf{NP}-complete if each
\textit{quintuplet} of consecutive vertices were a clique. This,
however, fails to be the case. A \textit{trilateration graph} in
$\mathbb{R}^D$ is a graph with an order $(v_1,\ldots,v_n)$ on the
vertices such for all vertices $v_i$ with $i>D+1$, $\{j,i\}\in E$ for
all $j\in\{i-D-2,\ldots,i-1\}$ (i.e.~each vertex is adjacent to the
preceding $D+1$ vertices). In three-dimensional space, this implies
having distances to at least 4 vertices earlier in the order, which
means having a clique for each consecutive quintuplet. By \cite{er04}
(Theorem~9), the MDGP problem associated to a trilateration graph can
be solved in polynomial time.

\section{Branch-and-Prune algorithm}
\label{bp}
In this section we present a Branch-and-Prune (BP) algorithm for the
DMDGP. The approach mimicks the structure of the problem closely: at
each step we can place the $i$-th atom in two possible positions
$x_{i},x'_{i}$. However, either or both of these positions may be
infeasible with respect to a number of constraints. The search is
branched on all atomic positions which are feasible with respect to
all constraints; by contrast, if a position is not feasible the search
scope is pruned. In this context we call the feasibility verifications
{\it pruning tests}. The simplest (but very effective) type of these
is the Direct Distance Feasibility (DDF) pruning tests: for all
distance pairs $\{j,i\}\in F$ (with $j<i-4$, see Sec.~\ref{intro},
p.~\pageref{EFH}) we check that $(||x_j - x_{i}||^2 -
d^2_{j,i})^2<\varepsilon$, where $\varepsilon>0$ is a given
tolerance. If the inequality does not hold, we prune the search node.

The BP algorithm is therefore an algorithmic framework whose
definition is completed by expliciting the pruning tests. These can be
of geometrical or of physical-chemical nature. These can be of
geometrical or of physical-chemical nature. Apart from the DDF pruning
tests, many other tests are possible; a few of them are discussed
below. This algorithm, as described, will find all solutions to the
problem. If we are interested in one solution only, we can stop the
search as soon as we have placed the last atom in a feasible position.

\subsection{Algorithmic Framework}
\label{algframe}
Let $T$ be a graph representation of the search tree. Initially, $T$
is initialized to the search nodes $1\rightarrow 2\rightarrow
3\rightarrow 4$ (no branching) since the first three atoms can be
fixed to feasible positions $x_1,x_2,x_3$ and the fourth atom $x_4$
can be fixed to any of its two possible positions by Theorem
\ref{symthm}. By the current rank of the search tree we mean the index
of the atom being placed at the current node. At each search tree node
of rank $i$ we store:
\begin{itemize}
\item the position $x_i\in\mathbb{R}^3$ of the $i$-th atom;
\item the cumulative product $Q_i = \prod_{j=1}^i B_j$ of the torsion
  matrices;
\item a pointer to the parent node $P(i)$;
\item pointers to the subnodes $L(i),R(i)$ (initialized to a dummy
  value $\mbox{PRUNED}$ if infeasible).
\end{itemize}
Notice that the edge structure of the graph $T$ is encoded in the
operators $P(),L(),R()$ defined at each node. The recursive procedure
at rank $i-1$ is given in Algorithm \ref{algbp}. Let
$y=\transpose{(0,0,0,1)}$, $\varepsilon>0$ a given tolerance and $v$ a
node with rank $i-1$ in the search tree $T$.
\begin{algorithm}[!ht]
\begin{algorithmic}
\item BranchAndPrune($T$, $v$, $i$)
\IF{($i\le n-1$)}
  \item {\sc Compute the possible placements for $i$-th atom}:
  \item calculate the torsion matrices $B_i,B'_i$ via
  Eq.~(\ref{bi});
  \item retrieve the cumulative torsion matrix $Q_{i-1}$ from the
  parent node $P(v)$;
  \item compute $Q_i=Q_{i-1}B_i$, $Q'_i=Q_{i-1}B'_i$ and $x_i,x'_i$ from
  $Q_iy,Q'_iy$;
  \item let $\lambda=1,\rho=1$;
  \item {\sc Pruning tests}: \label{pruningtest}
  \IF{($x_i$ is feasible)}
    \item create a node $z$, store $Q_i$ and $x_i$ in $z$, let
    $P(z)=v$ and $L(v)=z$;  
    \item set $T\leftarrow T\cup\{z\}$;
    \item BranchAndPrune($T$, $z$, $i+1$);
  \ELSE
    \item set $L(v)=\mbox{PRUNED}$;
  \ENDIF
  \IF{($x'_i$ is feasible)}
    \item create a node $z'$, store $Q_i$ and $x_i$ in $z'$, let
    $P(z)=v$ and $R(v)=z'$;
    \item set $T\leftarrow T\cup\{z'\}$;
    \item BranchAndPrune($T$, $z'$, $i+1$);
  \ELSE
    \item set $R(v)=\mbox{PRUNED}$;
  \ENDIF
\ELSE
  \item {\sc Rank $n$ reached, a solution was found}:
  \item solution stored in parent nodes ranked $n$ to 1, output by
  back-traversal; 
\ENDIF
\end{algorithmic}
\caption{BP algorithm.}
\label{algbp}
\end{algorithm}

\subsection{Pruning tests and error tolerance}
\label{correction}
There are two types of experimental errors arising from NMR distance
measurements: (i) systematic uncertainty on each measurement, and (ii)
a certain percentage of completely wrong measurements \cite{bkl99}.
Errors of the first type are usually dealt with by introducing
distance bounds \cite{mw99}, which the BP algorithm can take into
account without any problem. To the best of our knowledge, errors of
the second type have only been tackled by the Error Correcting Code
(ECC) proposed in \cite{bkl99}. Naturally, this ECC can (and should)
be applied to the protein backbone distance matrix as a preprocessing
step to running the BP algorithm. On top of this, however, many of the
pruning tests are ``natively'' suited for attempting to correct this
type of error probabilistically, if a measure of the infeasibility is
provided by the test. We only show here how to adapt the DDF tests for
the two types of NMR errors.

If we consider distance bounds like $d_{ji}^L\le d_{ji}\le d_{ji}^U$
for each $\{j,i\}\in F$ we simply have to modify the pruning tests as
follows. Placing atom $i$ at search node $v$, for
$j\in\bar{\delta}(i)\cap F$,
\begin{enumerate}
\item for $L(v)$: if $d^L_{ji}\le ||x_j-x_i|| \le
d^U_{ji}$ then $x_i$ is feasible, else it must be pruned;
\item for $R(v)$: if $d^L_{ji}\le ||x_j-x'_i|| \le
d^U_{ji}$ then $x'_i$ is feasible, else it must be pruned.
\end{enumerate}
As for the errors of the second type, let $100p$ (with $p\in[0,1]$) be
the known average percentage of completely wrong measurements. We deal
with these in a probabilistic way: suppose we are positioning the
$i$-th atom in position $x_i$, w.l.o.g.~at the left node $L(v)$ (the
reasoning for the right node is the same). A distance $d_{ji}$ is {\it
infeasible} for $x_i$ if the corresponding {\sc Pruning test} for
$L(v)$ fails. We prune $L(v)$ from the search tree only if more than
$p|\bar{\delta}(i)\cap F|$ distances are infeasible for $L(v)$. The
downside of this method is that it may introduce some false positives
in the solution set.

\subsection{Euclidean bounds pruning tests}
\label{eucbound}
These tests employ the fact that inter-atomic distances are assumed to
be Euclidean. Much like the pruning of the search scope in
point-to-point Dijkstra shortest-path searches on Euclidean graphs, we
can prune away an atomic position $i$ if it is too far with respect to
the given distances. Consider atoms $h,i,k$ with $h<i<k$ such that
$\{h,k\}\in E$ (so that $d_{hk}$ is known). Assume that the BP has
already placed atom $h$, and that we are now verifying feasibility for
atom $i$. Let $D(i,k)$ be an upper bound to the distance $||x_i-x_k||$
for all possible immersions $x:G\rightarrow\mathbb{R}^3$ which are
feasible DMDGP solutions.

\begin{lem}
If $D(i,k)<||x_h-x_i||-d_{hk}$ for all feasible
$x:G\rightarrow\mathbb{R}^3$, then the BP search node for atomic
position $x_i$ can be pruned.
\label{lemeuc}
\end{lem}
\begin{proof}
Suppose, to get a contradiction, that position $x_i$ is feasible for
the DMDGP instance being solved. By definition, $D(i,k)\ge
||x_i-x_k||$. Since distances are Euclidean, $||x_i-x_k||\ge
||x_h-x_i||-||x_h-x_k||$. Hence $D(i,k)\ge ||x_h-x_i||-d_{hk}>
D(i,k)$, which is a contradiction.
\end{proof}

By Prop.~\ref{lemeuc}, every upper bound $D(i,k)$ to the distance
$||x_i-x_k||$ provides a valid pruning test. Furthermore, in all
Euclidean graphs the Euclidean distance between two vertices is a
lower bound to the cost of all paths joining the two vertices in the
graph. We therefore let $D(i,k)$ be the cost of the shortest path from
$i$ to $k$ in $G$, which provides a valid pruning test.

\subsection{Continuous optimization-based pruning tests}
\label{contopt}
Global optimization techniques for the smooth nonconvex formulation
(\ref{continuous}) of the MDGP \cite{h95,mw97,mw99,lln1,doublevns} can
be employed to verify that the current backbone is feasible. As most
of these techniques are usually computationally expensive, this type
of pruning tests may only be performed once in a while. Although the GCA
\cite{mw97,mw99} scores the best computation times, it usually
provides solutions of rather low accuracy, so it may not be the best
candidate.

\subsection{Physical-chemical pruning tests}
\label{physchem}
As mentioned in the introduction, a SCPP can be solved for every
backbone in order to try to place the side-chain residues onto the
backbone. If the SCPP is infeasible, this means that the backbone is
also infeasible. Therefore, for every partial backbone we can try to
solve the associated SCPP to attempt to prune some search
branches. Since the SCPP is {\bf NP}-complete (hence its solution may
be computationally very expensive), this type of pruning tests should
only be performed once in a while. 

Other types of physical-chemical pruning tests can be devised on a
per-molecule basis.

\subsection{Detailed example}
\label{secex}
In this section we discuss the application of Algorithm \ref{algbp}
to a simple example (artificially generated as explained in
\cite{lavor}, also see Section \ref{seclavorinstances}). 

\begin{figure}[!ht]
\begin{center}
\fbox{\includegraphics[width=13cm]{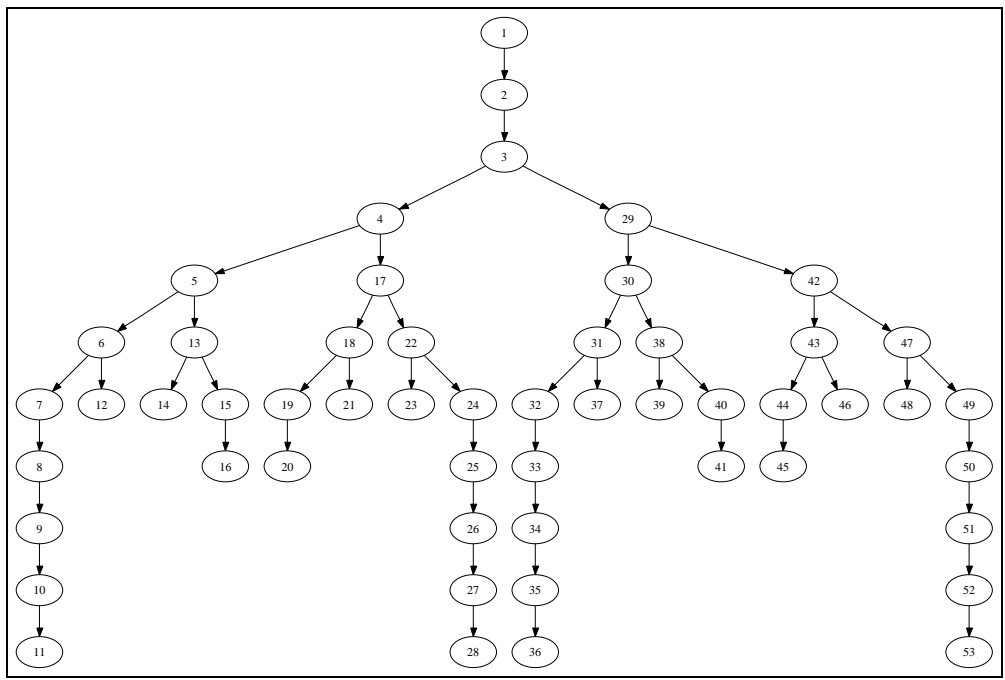}}
\end{center}
\caption{The BP tree of the instance of Section \ref{secex}.}
\label{figbpex}
\end{figure}

\begin{figure}[!ht]
\begin{center}
\fbox{\includegraphics[width=8cm]{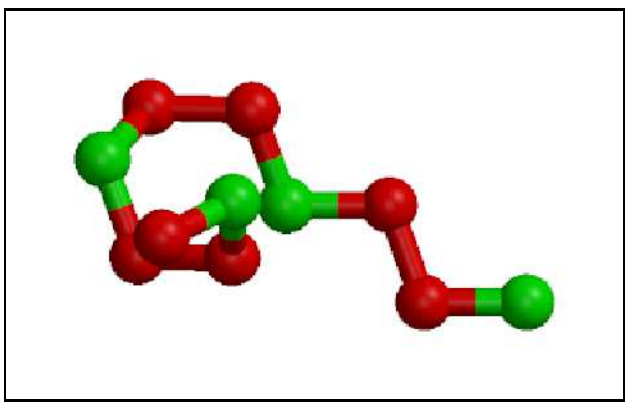}}
\end{center}
\caption{One of the possible solutions of the {\tt lavor11\_7} instance.}
\label{figlav117}
\end{figure}

The instance in question (called {\tt lavor11\_7}), with all bond
angles set to 1.91 radians, has 11 atoms:
\begin{eqnarray*}
\delta(2)\cup F &=& \{9\}, d^F_2=(3.387634917) \\
\delta(3)\cup F &=& \{8,9,10\}, d^F_3=(3.96678038,3.003368265,3.796280236) \\
\delta(4)\cup F &=& \{8,9,10\}, d^F_4=(2.60830758,2.102385055,3.159309539) \\
\delta(5)\cup F &=& \{9,10\}, d^F_5=(2.689078459,3.132251169) \\
\delta(6)\cup F &=& \{10\}, d^F_6=(3.557526815) \\
\delta(7)\cup F &=& \{11\}, d^F_7=(3.228657023).
\end{eqnarray*}
The distances in $H$ are of course $\delta(i)\cup H=\{i+1,i+2,i+3\}$
for all $i\le n-3$, $\delta(n-2)\cup H=\{n-1,n\}$, $\delta(n-1)\cup
H=\{n\}$. The vector of the distances in $H$ is:
\begin{eqnarray*}
d^H &=& (1.526,2.491389536,3.83929637,\\
    & &  1.526,2.491389536,3.831422399, \\ 
    & &  1.526,2.491389536,3.835602674, \\
    & &  1.526,2.491389535,3.030585263, \\
    & &  1.526,2.491389534,2.899348439, \\
    & &  1.526,2.491389535,3.086914764, \\
    & &  1.526,2.491389536,2.788611167, \\
    & &  1.526,2.491389536,2.888815709, \\
    & &  1.526,2.491389537,\\
    & &  1.526). 
\end{eqnarray*}
As can be seen from the BP tree given in
Fig.~\ref{figbpex} (this is actually the output of Algorithm
\ref{algbp} on the given instance), this instance has four solutions:
the leaf nodes at rank 11 --- the rank is given by the number of the
leftmost node in each row.
Notice that the earliest node when some pruning occurs is at rank 7,
i.e. no pruning occurs before the placement of the 8-th atom. This
happens because there are no distances $\{j,k\}\in F$ with $k<8$, so
each position for atoms with index $i<8$ is feasible (by construction
of $x_i,x'_i$) with the distances in $F$. Again, there is pruning at
ranks 8, 9, 10, i.e. during the placement of atoms 9, 10, 11, because
there are distances $\{j,k\}\in F$ with $k=9, 10, 11$. One of the
solutions is shown in Fig.~\ref{figlav117}.

\section{Computational Results}
\label{compres}
In order to test the viability of the proposed method, we tested a
class of randomly generated MDGP instances described in
\cite{lavor}. We present comparative results of BP (where only the DDF
pruning tests have been implemented) and another existing MDGP
software called {\tt dgsol} implementing the GCA \cite{mw99}. It turns
out that BP is always superior to {\tt dgsol} for solution accuracy,
generally superior as regards speed, and inferior as regards memory
requirements. It is fair to remark here that the GCA is able to solve
the MDGP, whereas BP is limited to solving the DMDGP only. In this
sense the present comparison is not completely fair to GCA.

\subsection{Software testbeds}
\label{secsw}
The software code {\tt dgsol} \cite{mw99} (version 1.3) can be freely
downloaded from 
\begin{quote}
{\tt http://www.mcs.anl.gov/\~{}more/dgsol/}.
\end{quote}
The algorithm implemented by the {\tt dgsol} code is very different
from ours. First, it targets a more general problem class: the
Molecular Distance Geometry Problem with Distance Bounds. In this
problem, lower and upper bounds to atomic distances are known, rather
than the exact atomic distances. Since these are usually estimated
through NMR techniques, it is realistic to assume that there is an
experimental error in the measurements (our approach does not consider
this issue yet). Secondly, {\tt dgsol} needs to make no assumption
whatsoever about the distances of triplets and quadruplets of
consecutive atoms being known.  Thirdly, {\tt dgsol} is based on a
continuous smoothing of the original problem to a form which has fewer
local minima. An ordinary NLP optimization method is then applied to
the modified problem, and the optimum is traced back to the original
problem. This is a fully continuous optimization algorithm, whereas BP
is a discrete method.

It turns out that the main advantages of BP over {\tt dgsol} are:
\begin{enumerate}
\item tractability of larger instances;
\item higher solution accuracy;
\item BP can potentially find {\it all} feasible solutions, not just one.
\end{enumerate}
By contrast, the main advantages of {\tt dgsol} over BP are:
\begin{enumerate}
\item it targets a larger class of problems;
\item its running time seems to increase very slowly (and regularly)
  as a function of the number of atoms in the molecule, at least when
  the set of given distances is comparatively small;
\item the amount of memory needed to complete the search is negligible.
\end{enumerate}
The BP algorithm behaves very unpredictably with respect to the amount
of needed memory, sometimes requiring over 1GB RAM for relatively
small molecules (40 atoms), sometimes solving 1000-atoms instances in
a few seconds and very little memory. 

\subsection{Mor\'e-Wu instances}
\label{secmorewuinstances}
\begin{figure}[!ht]
\psfrag{x}{$x$}
\psfrag{y}{$y$}
\psfrag{z}{$z$}
\begin{center}
\fbox{\includegraphics[width=7cm]{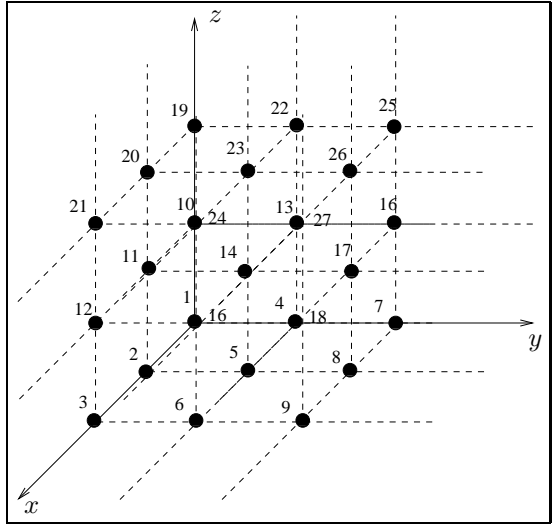}}
\end{center}
\caption{The $s=3$ Mor\'e-Wu instance with 27 atoms.}
\label{figmorewu}
\end{figure}

The Mor\'e-Wu instances are finite three-dimensional hypercubic
lattices with $s^3$ atoms, as shown in Fig.~\ref{figmorewu}. The bond
lengths parallel to the coordinate axes are assumed to be 1.
By providing the instances with the obvious atomic ordering (as shown
in Fig.~\ref{figmorewu}) and the bond angles, we can make them
amenable to the application of our method. However, since many of the
bond angles $\vartheta$ are equal to $\pi$ (e.g. the angle between
atom 1 and atom 3 in Fig.~\ref{figmorewu}), these instances are
undiscretizable (see Sect.~\ref{secundiscr}). In particular we get
$\sin\vartheta=0$, so Eq.~(\ref{pog}) ceases to hold.

\begin{figure}[!ht]
\psfrag{x}{$x$}
\psfrag{y}{$y$}
\psfrag{z}{$z$}
\begin{center}
\fbox{\includegraphics[width=7cm]{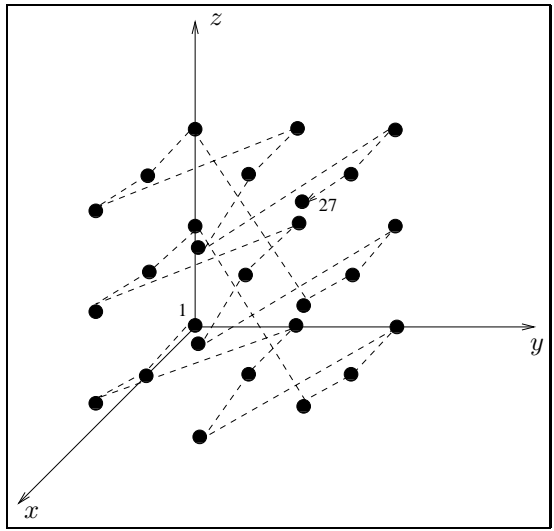}}
\end{center}
\caption{The modified Mor\'e-Wu instance with 27 atoms. The dashed
  arrow indicates the atomic ordering.}
\label{figmodmorewu}
\end{figure}

In order to test these instances, we perturbed the lattice points
$x_i=(x_{i1},x_{i2},x_{i3})$ in the following way:
\begin{equation}
  \forall i\le s^3 \; (i\!\!\mod 3 = 0 \Rightarrow (x_{i3} \leftarrow
  x_{i3} + (-1)^i\eta)),
\end{equation}
where $\eta$ was taken to be 0.25. This gave rise to instances which
we call ``modified Mor\'e-Wu instances'' ({\tt mmorewu-}$s$, where
$s^3$ is the number of atoms in the molecule). An example is shown in
Fig.~\ref{figmodmorewu}. It is worth mentioning that as the original
Mor\'e-Wu instances describe a molecular structure rarely, if ever,
found in proteins, we feel our perturbation does not alter crucial
molecular characteristics.
We generated all modified Mor\'e-Wu instances from $s=2$ to
$s=6$.

\subsection{Lavor instances}
\label{seclavorinstances}
These instances, described in \cite{lavor}, are based on the model
proposed by \cite{prw96}, whereby a molecule is represented as a
linear chain of atoms. Bond lengths and angles are kept fixed, and a
set of likely torsion angles is generated randomly. Depending on the
initial choice of bond lengths and angles, the Lavor instances give
rather more realistic models of proteins than other randomly generated
instances do (see for example the instances described in
\cite{mw99}). Fig.~\ref{figlav117} gives an example of a Lavor
instance. In the numerical tables, we labelled the Lavor instances by
{\tt lavor}$n$-$m$, where $n$ is the number of atoms in the molecule
and $m$ is an instance ID (since there is a random element of choice
in the generation of the Lavor instances, many different instances can
be generated having the same atomic size).

We generated 10 different Lavor instances for each size
$n=10,\ldots,70$ (``small set''), and 4 different Lavor instances for
each size $n$ in $\{100i | 1\le i\le 10\}$ (``large set'').

\subsection{Hardware and memory considerations}
\label{secmemory}
All tests have been carried out on an Intel Pentium IV 2.66GHz with
1GB RAM, running Linux. The code implementing the BP algorithm has
been compiled by the GNU C++ compiler v.3.2 with the {\tt -O2} flag.
As mentioned above, BP can be very memory-demanding. We deliberately
took the choice of employing a low-end PC with just 1GB RAM to show
just how powerful this technique can be even with modest hardware.

The BP algorithm is in general very fast, since all it does is testing
feasibility with the known distances at each branched node. However,
exploring the search space may require a lot of memory, especially if
no pruning occurs early in the run. Consequently, when the physical
RAM of the test machine is exhausted, and the operating system starts
swapping to disk, the total CPU elapsed time size becomes
unmanageable. Thus, it was decided to kill all jobs requiring more
than 1 GB RAM. In particular, we solved almost all the Lavor instances
in the ``small set'' and found one solution for each of the Lavor
instances in the ``large set''.

\subsection{Comparative results}
\label{seccompres}
The full results table for the complete test suite includes 655
instances and spans 14 pages: thus, only a sample will be presented in
detail. The averages, however, are taken with respect to the whole
suite. The $\varepsilon$ parameter in the DDF pruning tests was set to
$1\times 10^{-3}$ for all tests. Table~\ref{tabcompres1} contains
detailed results for the sample. The instances are described by their
name, their atomic size $n$ and the number of given distances
$|S|$. Note that in order to use {\tt dgsol}, the lower and upper
bounds to these distances were set to $\pm 5\times 10^{-4}$. Other
than this, {\tt dgsol} was used with all default parameter values. The
results refer to three methods: {\tt dgsol}, BP stopped after the
first solution was found (BP-One), and BP run to completion
(BP-All). For {\tt dgsol} and BP-One, the user CPU time (in seconds)
was reported, as well as the Largest Distance Error (LDE), defined as
\begin{equation*}
  \mbox{LDE} = \frac{1}{|S|}\sum_{(i,j)\in S} \frac{| \; ||x_i-x_j|| -
  d_{ij} |}{d_{ij}},
\end{equation*}
employed as a measure of solution accuracy (the lower, the
better). For the (BP-All) method, we reported the user CPU time and
the number of solutions found (\#Sol). Missing values are due to
excessive memory requirements (over 1GB RAM).

\begin{table}[!ht]
\begin{center}
\begin{tabular}{|l|c|c||c|c||c|c||c|c|} \hline
\multicolumn{3}{|c||}{Instance} & \multicolumn{2}{c||}{\tt dgsol} & \multicolumn{2}{c||}{BP-One} & \multicolumn{2}{c|}{BP-All} \\ \hline
{\bf Name} & $n$ & $|S|$ & {\bf CPU} & {\bf LDE} & {\bf CPU} & {\bf LDE} & {\bf CPU} & {\bf \#Sol} \\ \hline 
{\tt mmorewu-2 }  &   8 &   28 &   0.02 &2.63E+5& 0.00 & 4.37E-10 & 0.00 & 2    \\
{\tt mmorewu-3 }  &  27 &  331 &   0.23 &  6.99 & 0.00 & 2.97E-09 & 0.00 & 2    \\
{\tt mmorewu-4 }  &  64 & 1882 &   0.67 &7.79E-6& 0.00 & 5.56E-09 & 0.00 & 4    \\
{\tt mmorewu-5 }  & 125 & 7105 &   2.94 &3.54E-6& 0.00 & 1.67E-08 & 0.01 & 4    \\
{\tt mmorewu-6 }  & 216 &21461 &  18.65 & 0.032 & 0.02 & 4.91E-08 & 0.03 & 4    \\
\hline
{\tt lavor10\_0 } &  10 &   33 &   0.02 &1.57E-5& 0.00 & 5.36E-10 & 0.00 & 4    \\
{\tt lavor15\_0 } &  15 &   57 &   0.10 &4.04E-5& 0.00 & 2.84E-09 & 0.00 & 16   \\
{\tt lavor20\_0 } &  20 &  105 &   0.14 &2.77E-5& 0.00 & 6.13E-09 & 0.00 & 8    \\
{\tt lavor25\_0 } &  25 &  131 &   0.84 &1.18E-4& 0.00 & 1.38E-09 & 0.00 & 8    \\
{\tt lavor30\_0 } &  30 &  169 &   0.40 &1.75E-5& 0.00 & 1.23E-09 & 0.00 & 2    \\
{\tt lavor35\_0 } &  35 &  171 &   0.81 &9.33E-5& 0.00 & 1.52E-09 & 0.00 & 64   \\
{\tt lavor40\_0 } &  40 &  295 &   2.84 & 0.096 & 0.00 & 2.87E-09 & 0.00 & 2    \\
{\tt lavor45\_0 } &  45 &  239 &   3.33 & 0.170 & 0.00 & 6.92E-09 & 0.00 & 2    \\
{\tt lavor50\_0 } &  50 &  271 &   3.45 & 0.696 & 0.00 & 3.96E-08 & 0.46 & 4096 \\
{\tt lavor55\_0 } &  55 &  551 &   5.80 & 0.257 & 0.00 & 2.66E-09 & 0.00 & 64   \\
{\tt lavor60\_0 } &  60 &  377 &   5.15 & 0.049 & 0.00 & 3.51E-09 & 0.00 & 64   \\
{\tt lavor65\_0 } &  65 &  267 &   2.61 & 0.065 & 0.00 & 7.76E-10 & --   & --   \\
{\tt lavor70\_0 } &  70 &  431 &   8.73 & 0.107 & 0.02 & 1.64E-09 & --   & --   \\
\hline
{\tt lavor100\_2} & 100 &  605 &   6.95 & 0.167 & 2.26 & 4.01E-09 & --   & --   \\
{\tt lavor200\_2} & 200 & 1844 &  63.52 & 0.395 & 0.00 & 5.66E-08 & --   & --   \\
{\tt lavor300\_2} & 300 & 2505 & 100.99 & 0.261 & 0.03 & 1.56E-08 & --   & --   \\
{\tt lavor400\_2} & 400 & 2600 & 182.21 & 0.767 & 0.01 & 3.35E-09 & --   & --   \\
{\tt lavor500\_2} & 500 & 4577 & 329.29 & 0.830 & 0.27 & 4.69E-07 & --   & --   \\
{\tt lavor600\_2} & 600 & 5473 & 299.76 & 0.700 & 0.01 & 4.94E-08 & --   & --   \\
{\tt lavor700\_2} & 700 & 4188 & 281.34 & 0.569 & 0.16 & 1.83E-06 & --   & --   \\
{\tt lavor800\_2} & 800 & 6850 & 570.20 & 0.528 & 3.34 & 3.37E-06 & --   & --   \\
{\tt lavor900\_2} & 900 & 7965 & 550.26 & 0.549 & 3.08 & 5.62E-06 & --   & --   \\
{\tt lavor1000\_2}& 1000& 8229 & 844.52 & 0.695 & 0.81 & 2.04E-06 & --   & --   \\
\hline
\end{tabular}
\end{center}
\caption{Computational results. Missing values are due to excessive
memory requirements ($>1$GB RAM).}
\label{tabcompres1}
\end{table}

It is immediately noticeable that whereas {\tt dgsol} always finds a
solution, BP sometimes fails to find one within 1 GB RAM. It is
instructive, however, to look at the solution accuracy (taken over the
whole test suite): whereas {\tt dgsol} ranges from $4.5\times 10^{-7}$
to $0.875$ (excepting a couple of out-of-scale values clearly due to
some numerical instability), BP scores a rather more impressive 
$4.74\times 10^{-11}$ to $5.62\time 10^{-6}$. On average, the solution
accuracy obtained by {\tt dgsol} is $9.55\times 10^{-2}$ whereas BP
averages $4.56\times 10^{-8}$. Furthermore, all the instances in the
Lavor ``large set'' are solved by {\tt dgsol} to a solution accuracy
of order $10^{-1}$: given that in BP pruning often occurs for
feasibility differences of order $10^{-1}$ and even $10^{-2}$, such a
slack solution accuracy may mean that {\tt dgsol} is not actually
finding the correct solution.

\begin{figure}[!ht]
\begin{center}
\includegraphics[width=10cm]{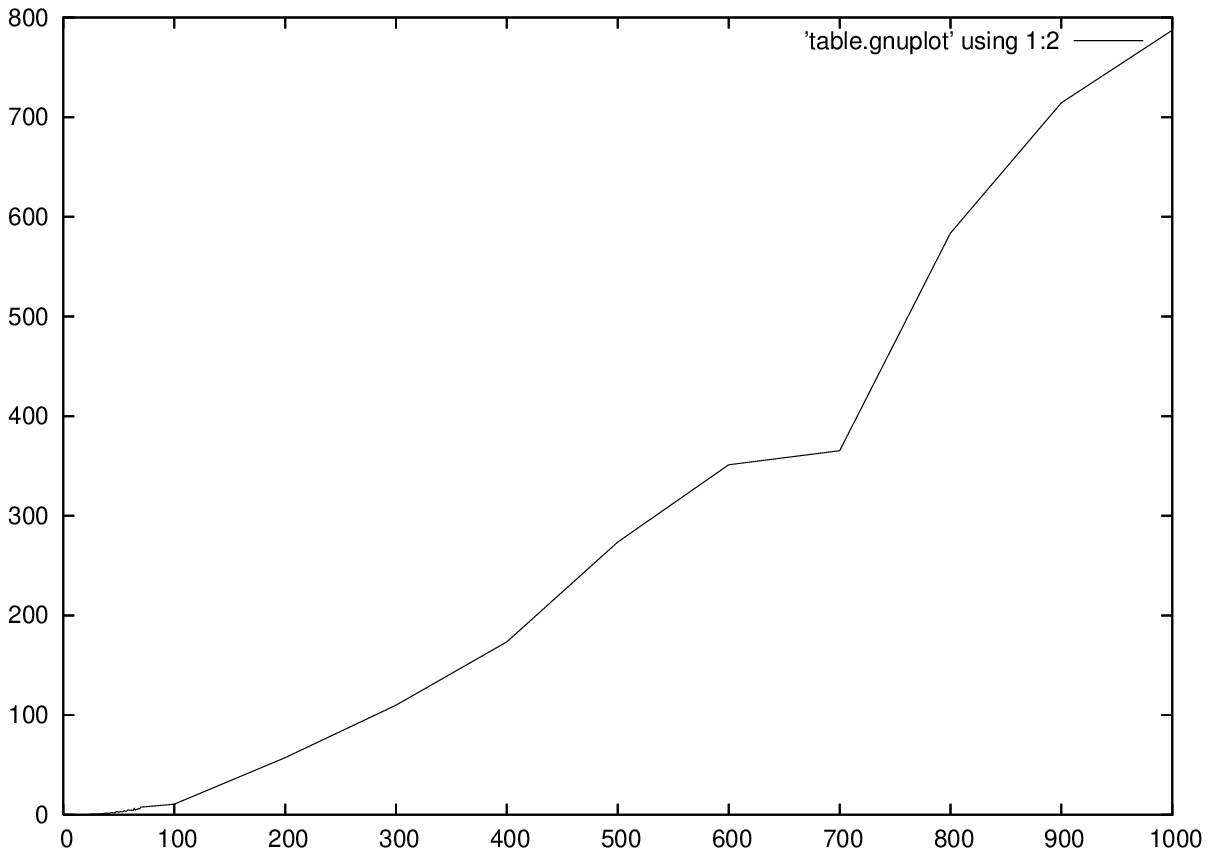}
\includegraphics[width=10cm]{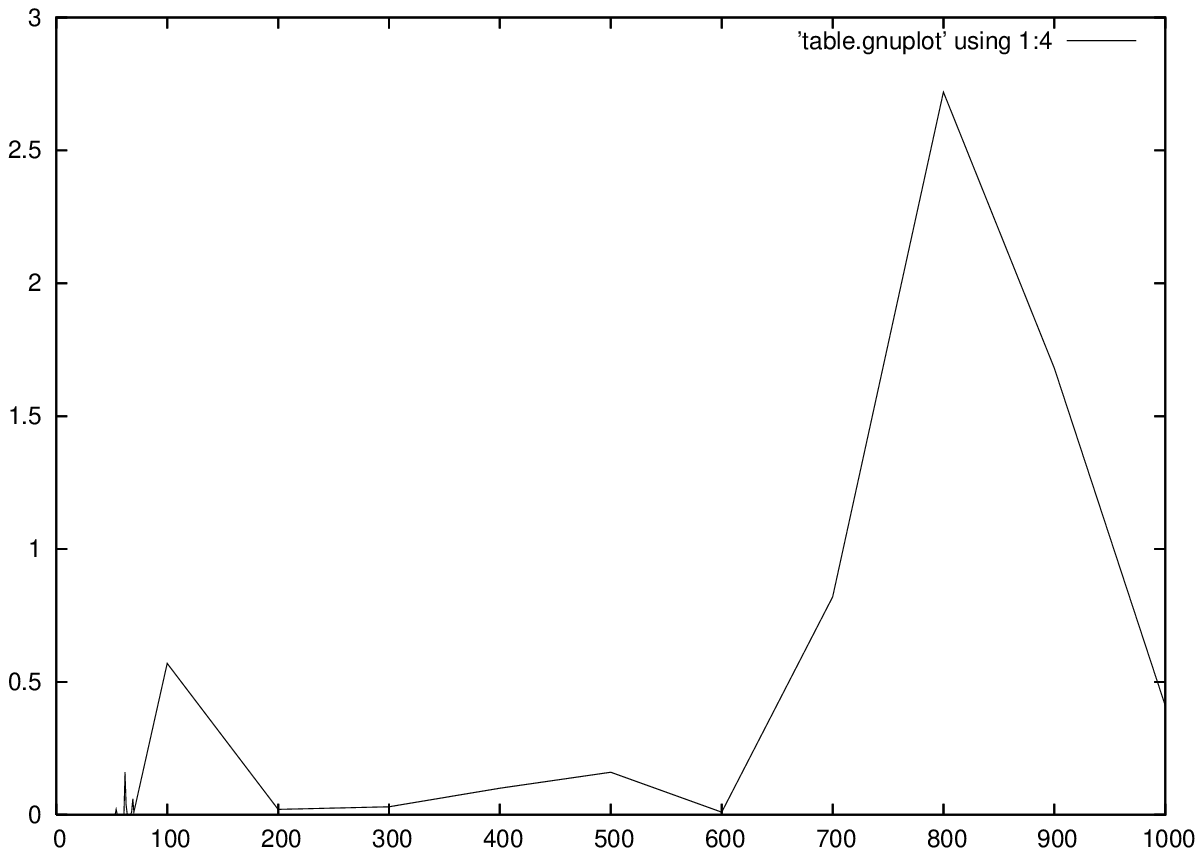}
\end{center}
\caption{Average user CPU time (plotted against molecular size) taken
  by {\tt dgsol} (top) and BP-One (bottom).}
\label{figcpuavg}
\end{figure}

\begin{figure}[!ht]
\begin{center}
\includegraphics[width=10cm]{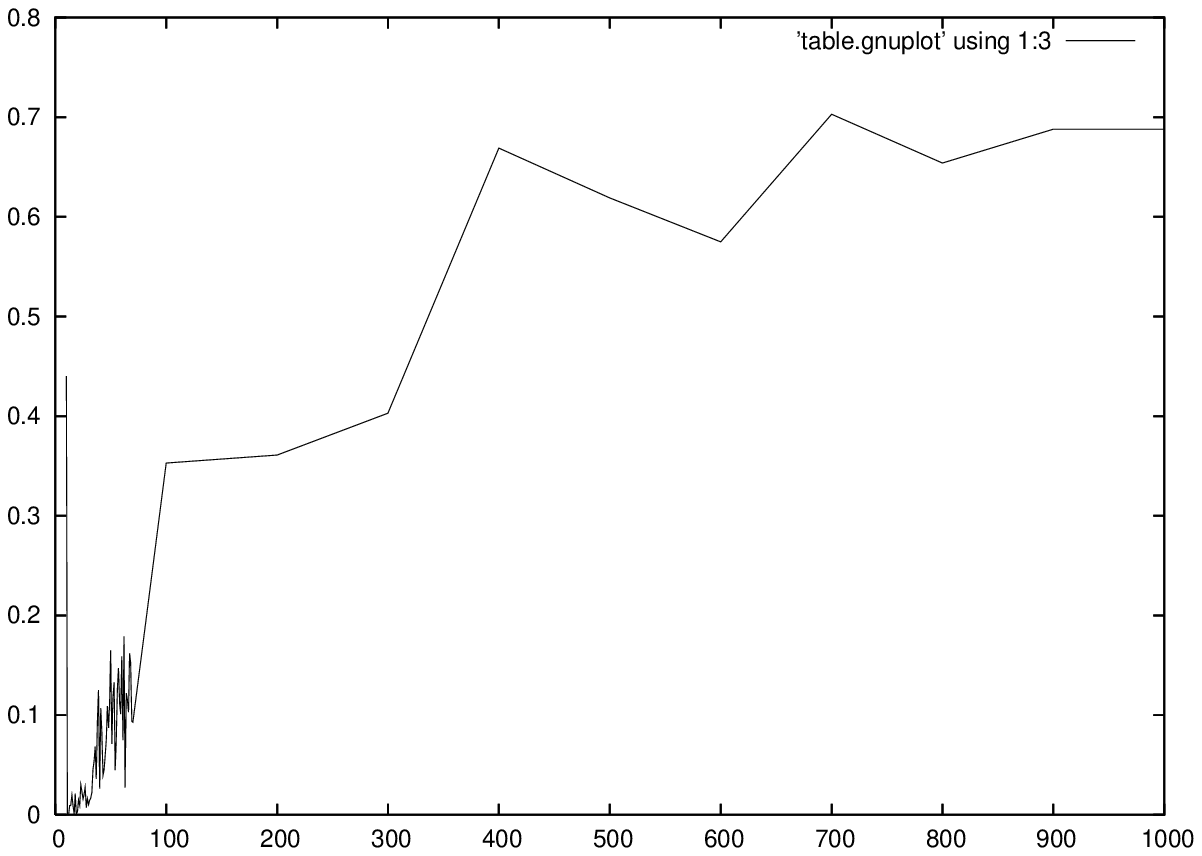}
\includegraphics[width=10cm]{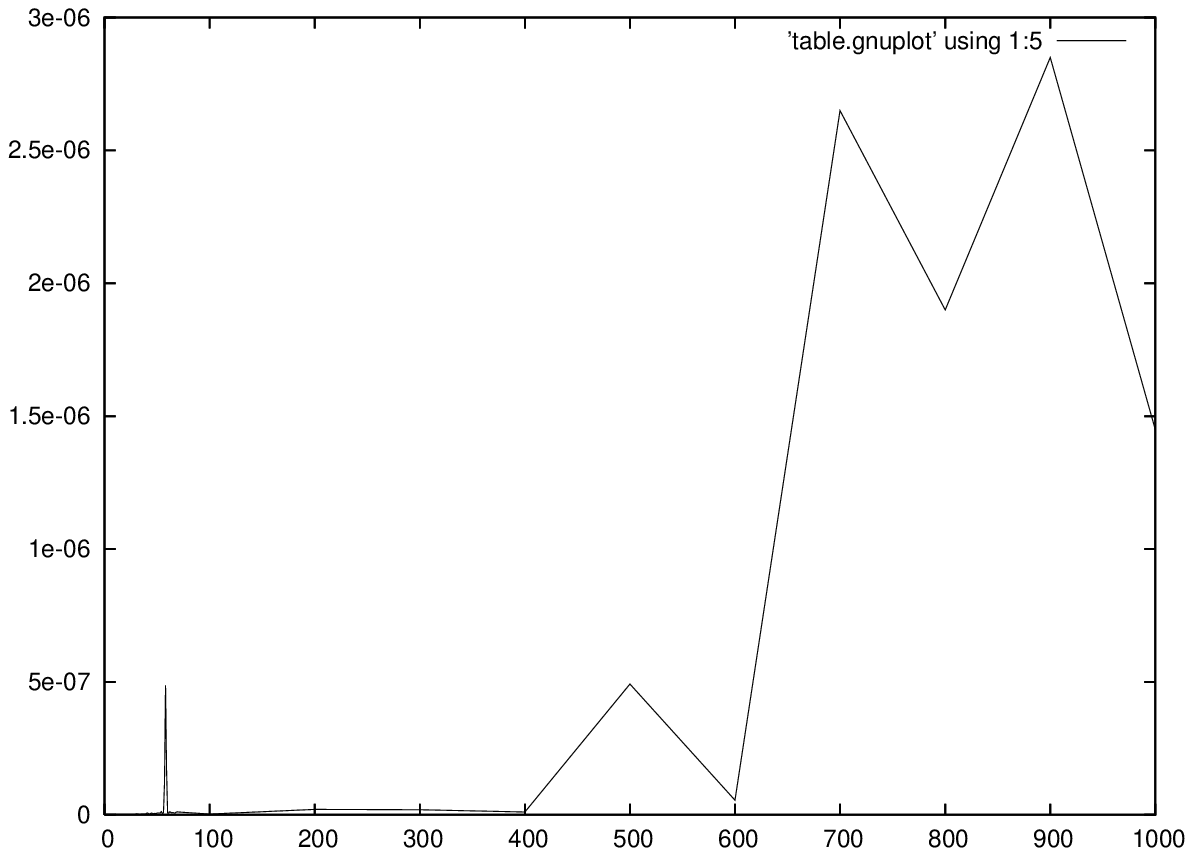}
\end{center}
\caption{Average accuracy (plotted against molecular size) attained
  by {\tt dgsol} (top) and BP-One (bottom).}
\label{figldeavg}
\end{figure}

Table~\ref{tabavg} reports the averages of the same parameters as in
Table~\ref{tabcompres1} taken over 10 Lavor instances in a sample of
the ``small set'' and over 4 Lavor instances in a sample of the
``large set''. It appears clear from these data that BP's strong
points are indeed speed and accuracy. A graphical representation of
the averages taken over the whole Lavor test set is shown in
Fig.~\ref{figcpuavg} (user CPU average taken to solve the instances in
function of the molecular size by {\tt dgsol} and BP-One) and
Fig.~\ref{figldeavg} (average accuracy of the solution attained by
{\tt dgsol} and BP-One). Notice the huge $y$-axis scale difference in
the two pairs of plots (around 300 times smaller in favour of BP for
CPU and around 30000 times smaller in favour of BP for accuracy).

\begin{table}[!ht]
\begin{center}
\begin{tabular}{|c||c|c||c|c||c|c|} \hline
\multicolumn{1}{|c||}{Instance} & \multicolumn{2}{c||}{\tt dgsol / avg.} & \multicolumn{2}{c||}{BP-One / avg.} & \multicolumn{2}{c|}{BP-All / avg.} \\ \hline
$n$ & {\bf CPU} & {\bf LDE} & {\bf CPU} & {\bf LDE} & {\bf CPU} & {\bf \#Sol} \\ \hline 
10 & 0.03 & 4.40E-01 & 0.00 & 1.19E-09 & 0.00 & 1.54E+01 \\
15 & 0.08 & 1.96E-02 & 0.00 & 1.23E-09 & 0.00 & 3.72E+01 \\
20 & 0.23 & 3.20E-03 & 0.00 & 1.94E-09 & 0.00 & 6.90E+01 \\
25 & 0.56 & 1.58E-02 & 0.00 & 1.58E-09 & 0.02 & 1.14E+02 \\
30 & 0.65 & 1.03E-02 & 0.00 & 3.45E-09 & 0.01 & 2.65E+02 \\
35 & 1.10 & 5.43E-02 & 0.00 & 2.84E-09 & 0.10 & 3.35E+03 \\
40 & 1.41 & 2.61E-02 & 0.00 & 5.75E-09 & 0.02 & 8.48E+02 \\
45 & 2.13 & 5.80E-02 & 0.00 & 6.25E-09 & 0.12 & 2.48E+03 \\
50 & 2.54 & 1.65E-01 & 0.00 & 6.62E-09 & 0.16 & 1.80E+03 \\
55 & 4.10 & 7.29E-02 & 0.00 & 5.53E-09 & 0.03 & 4.28E+02 \\
60 & 4.47 & 1.59E-01 & 0.00 & 6.44E-09 & 0.04 & 3.49E+02 \\
65 & 4.64 & 1.16E-01 & 0.00 & 8.37E-09 & 1.21 & 3.80E+03 \\
70 & 7.63 & 9.28E-02 & 0.01 & 1.07E-08 & -- & --      \\ \hline
100 & 10.57 & 3.53E-01 & 0.57 & 2.46E-09 & --   & --          \\
200 & 57.34 & 3.61E-01 & 0.02 & 2.00E-08 & --   & --          \\
300 & 109.91 & 4.03E-01 & 0.03 & 1.90E-08 & --   & --         \\
400 & 173.54 & 6.69E-01 & 0.10 & 1.05E-08 & --   & --         \\
500 & 273.66 & 6.19E-01 & 0.16 & 4.92E-07 & --   & --         \\
600 & 351.15 & 5.75E-01 & 0.01 & 5.47E-08 & --   & --         \\
700 & 365.37 & 7.03E-01 & 0.82 & 2.65E-06 & --   & --         \\
800 & 583.65 & 6.54E-01 & 2.72 & 1.90E-06 & --   & --         \\
900 & 714.39 & 6.88E-01 & 1.68 & 2.85E-06 & --   & --         \\
1000 & 787.30 & 6.88E-01 & 0.41 & 1.45E-06 & --   & --        \\
\hline
\end{tabular}
\end{center}
\caption{Average statistics for Lavor instances (over 10 instances for
  the set of small instances and over 4 for the set of large
  instances).}
\label{tabavg}
\end{table}

\subsection{The number of solutions}
It is remarkable that in Table~\ref{tabcompres1} BP-All always finds a
number of solutions which is a power of 2. Although we were not able
to ascertain the exact reason why the Mor\'e-Wu or Lavor instances had
these properties, we were able, through Theorem~\ref{complexitythm},
to ascertain that this behaviour does not apply to all instances in
the DMDGP.

\begin{lem}
The instance $I=\{101,102,104,108,1001,1002,1004,1008\}$ to the 
{\sc Subset-Sum} problem has exactly 3 solutions.
\end{lem}
\begin{proof}
We denote by $S,\bar{S}$ the partition of $I$ solving the {\sc
Subset-Sum} problem. Let 1008 be in $S$, then exactly one of
$\{1001,1002,1004\}$ must also be in $S$, and the other two in
$\bar{S}$ This force set membership of 101, 102, 104, 108 in the
following way: if $1000+x$ is in $S$, then $100+x$ is in $\bar{S}$ and
vice versa, for all $x\in\{1,2,4,8\}$.
\end{proof}

By the above Lemma and the proof of Theorem~\ref{complexitythm}, there
is a DMDGP instance with $2\times 27$ solutions (the 2 factor is due
to Theorem~\ref{symthm}).

\section{Final Remarks}
\label{conclusion}
In this paper we formally define an ${\bf NP}$-complete subclass of
the Molecular Distance Geometry Problem, related to proteins, for
which a discrete formulation can be supplied. Instances of this class
can be solved by employing a Branch-and-Prune algorithm which makes it
possible to find very efficiently one or all the solutions to the
problem instance. Furthermore, typical NMR measurement errors can be
taken into account by the algorithm. We illustrate the performance of
our algorithm on a set of randomly generated instances.

\section*{Acknowledgments}
The authors are profoundly indebted to Dr. C.~D\"{u}rr, to whom much
of the material on complexity is due. The authors would also like to
thank CNPq and FAPESP for their financial support.

\end{document}